\RequirePackage{etoolbox}

\documentclass[sn-mathphys]{sn-jnl}

\usepackage{tabularx}
\usepackage{booktabs}
\usepackage{array} 
\usepackage{longtable} 
\usepackage{subcaption} 
\usepackage{libertinus} 
\usepackage{placeins} 
\usepackage{needspace}

\jyear{2026}

\theoremstyle{thmstyleone}%

\theoremstyle{thmstyletwo}%
\theoremstyle{thmstylethree}%
\newtheorem{finding}{Finding}%

\usepackage{mdframed}
\mdfdefinestyle{findingbox}{%
  linecolor=black!25,
  linewidth=0.8pt,
  backgroundcolor=black!2,
  roundcorner=3pt,
  innertopmargin=-0.8\baselineskip,   
  innerbottommargin=0.6\baselineskip,
  innerleftmargin=0.9em,
  innerrightmargin=0.9em,
  skipabove=0.8\baselineskip,
  skipbelow=0.8\baselineskip
}
\surroundwithmdframed[style=findingbox]{finding}

\begin{document}

\title[Projecting the Emerging Mindset of SWE Agent]{Projecting the Emerging Mindset of SWE Agent by Launching a Wild Code Understanding Journey}

\author[1]{\fnm{Zhengyi} \sur{Zhuo}}
\author*[1]{\fnm{Yan} \sur{Liu}}\email{yanliu.sse@tongji.edu.cn}

\affil[1]{\orgdiv{School of Computer Science and Technology}, \orgname{Tongji University}, \orgaddress{\city{Shanghai}, \country{People's Republic of China}}}

\abstract{
Software engineering agents (SWE agents) are rapidly becoming popular in practice, yet we still know very little, in concrete, observable terms, about the behavioral regularities they exhibit when working in real codebases. Software engineering agents generate rich trajectories of tool use, intermediate reasoning, and self-directed stopping. This creates a basic contradiction: we know little about agent behavior, but we can collect abundant digital traces of it. The contradiction arises because trajectories record what the agent did, but not, by themselves, why those moves were chosen, what evidence was trusted, or what counted as sufficient understanding to stop. Still, precisely because the traces are faithful, replayable, and scalable, they offer one of the best empirical substrates for probing behavior and projecting the mindset underlying an agent's reasoning and choices.

To turn faithful traces into observable behavioral evidence, we build Ada as a scoped apparatus for repository-level code understanding and give it a world to move through. That world consists of real codebases made traversable through a bounded tool interface, so Ada’s exploration can remain open-ended in practice while becoming recordable as a finite trajectory. Inside this wild-but-bounded setting, Ada chooses where to look, what evidence to read closely, when to consolidate partial understanding, and when to close its account of the repository.

These journeys produce think-action chains in which the progression from navigation through evidence selection and synthesis to stopping becomes visible as situated behavior. We project these chains through carefully designed lenses that offer usable distances for reading agentic behaviors, without reducing interpretations to a mechanistic accounting of tool calls or inflating them into speculative overreading of hidden intent. Read together, the views enable a nascent projection of the SWE-agent mindset, expressed as behavioral profiles grounded in the agent’s recorded movement through real software worlds.

Together, the apparatus, lenses, and 408-trajectory study provide a methodological foundation for observing SWE-agent behavior in real codebases, showing how faithful digital traces can be transformed into disciplined, comparable projections of emerging agentic mindset. The resulting observations align with broader findings on agent efficiency, trajectory diversity, epistemic grounding, and the limits of intervention, giving additional support to the proposed method beyond the study's own experimental scope.

}


\maketitle

\section{Introduction}\label{sec:introduction}

Autonomous agents, boosted by LLM reasoning engines, are moving fast into the open world, increasingly operating through trajectories of tool use, environment interaction, intermediate reasoning, and self-directed stopping. A valid, useful agentic interaction is no longer a single-turn final answer. It usually unfolds as a group of ``turns'' through a world, cascading the actions and reasoning into a continuous, visible flow. From the flow, every thought and action is seemingly easier to read closely, providing a translucent view of agent behavior. 
However, agent behavior does not automatically become legible from the thought--action chain. Recorded trajectories do not, by themselves, reveal the agent's goals or strategic choices. Interpreting behavioral signals from outside the agent remains difficult, even though tools for logging and replaying agentic trajectories are mature and widely available \cite{liu2025graphectory, desmondt2025trajectoryexp, ou2025agentdiagnose}.

Software engineering agents (SWE agents) make this challenge unusually concrete. Their world is digital by nature, organized into repositories whose structure spans files and symbols, cross-file dependencies, and executable behavior, all inspectable as the agent moves through them. It is open enough to require genuine navigation, yet bounded enough for the resulting work to be recorded and revisited. A high degree of digitization, however, does not by itself guarantee an environment that is legible from outside the agent. When an SWE agent enters a repository, it must form a working picture of a software system from partial observations. It must decide where to look, what to trust, when to consolidate, and when its understanding is sufficient to stop. These trajectories offer a rare site for studying agentic work as situated movement through a structured world.

Interpreting the behavioral phenomena of an intelligent agent is pulled between two failure modes. At one extreme, a close reading of tool calls yields mostly mechanical signals rather than evidence of intelligent behavior; at the other, treating every move as motive imputes intention to every trace and therefore explains nothing. Still, an agent's ``mindset'' is not beyond reach. Between these extremes lies a workable stance in which behavioral interpretations remain accountable to the traces that generated them. In this work, we develop a reproducible methodology for studying SWE-agent behavior from an analytical distance where recorded trajectories can begin to reveal an emerging mindset.

Before any changes are made to the codebase, code understanding is the phase in which SWE agents (and also human developers) align their perceptions with the software system as it exists. This phase is especially revealing of mindset, because it is not reducible to task success: the agent must build understanding from partial evidence and then judge for itself when that understanding is sufficient to stop.

A different line of work looks inside the model itself. Mechanistic interpretability, advanced most visibly by Anthropic, attempts to recover features and circuits from internal activations, offering a glass-box account of what the network represents. This access is real, but it depends on substantial interpretability infrastructure and compute. It also characterizes the foundation model more directly than the agent assembled around it.

Our work starts from a complementary premise: even without opening the model, an agent's ``mindset'' leaves systematic traces in its thought--action chain. Reading mindset from trajectories, therefore, offers a distinct scientific stance. Because it reads the assembled agent rather than the foundation model, this stance is tied to the artifact that developers actually ship. It is also comparatively inexpensive and suitable for comparison across models and conditions. At the same time, the limits of this outside view are not obvious. We treat the observability of mindset from traces as an open question, so our method probes the extent to which such readings can be inferred.

\begin{figure}
    \centering
    \includegraphics[width=0.6\linewidth]{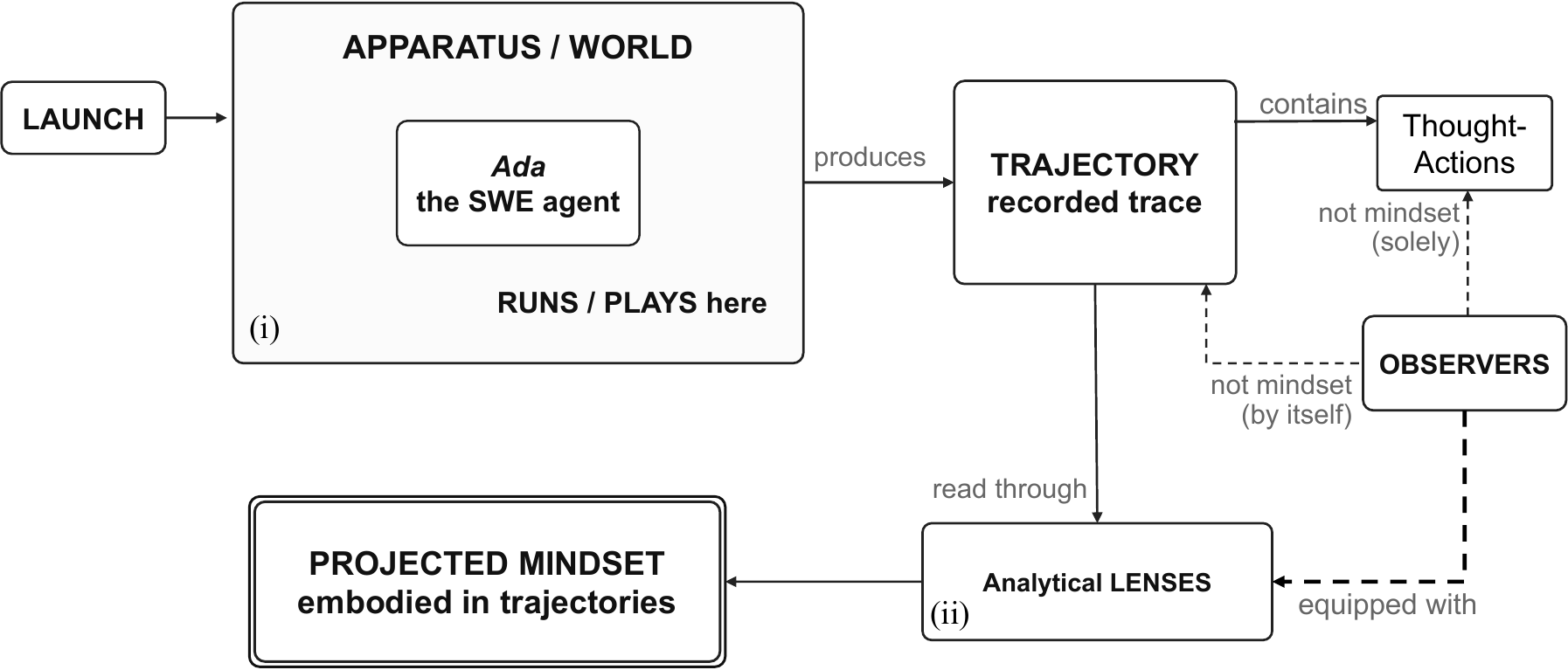}
    \caption{Overview of the outside-in reading approach: an SWE agent interacts with a repository world through a tool-mediated apparatus, producing a recorded trajectory that is projected through observation lenses into comparable behavioral signals.}
    \label{fig:meta-model}
\end{figure}

As illustrated in Fig.~\ref{fig:meta-model}, projecting mindset from outside requires (i) an environment that preserves genuine exploratory behavior while remaining bounded enough for repeated observation, (ii) analytical lenses that project recorded trajectories into comparable, interpretable views of how the agent oriented itself. This paper provides both. We build a minimal SWE agent apparatus that runs a multi-turn tool-use loop across six real-world repositories. A suite of observation lenses then renders different aspects of each trajectory into readable form. We use a dedicated code-understanding agent, \textbf{Ada}\footnote{Our code understanding SWE agent working inside the apparatus. Named after Ada Lovelace}, as the subject of study.

A recorded trajectory is not a direct readout of the agent's internal mindset. It provides a faithful account of observable choices: where the agent looked, what it treated as evidence, how it consolidated partial understanding, when it decided to stop. Our method uses lenses to project those choices into a scoped behavioral profile. The projection supports disciplined comparison, without claiming access to hidden intent.

Our contributions are as follows.

\begin{enumerate}[1.]
    \item Within repository-level code understanding, we construct a bounded-yet-wild software world with scoped task activators. This setting makes the agent's exploration recordable as trajectories that can be interpreted against the repository's independently known structure. The world is designed as an observational instrument, providing a stable ground for making agent behavior legible and comparable.
    \item  At the level of trajectory interpretation, we formalize a suite of observation lenses that project recorded thought--action chains into structured views. These lenses provide usable analytical distances, from aggregate activity to temporal shape, action transitions, semantic grounding, and repository contact, without collapsing trajectory interpretation into raw tool-call accounting or superficial speculations.
    \item For probing mindset observability, we operationalize a controlled launch-and-intervention framework. Prompt-framing conditions expose how exploration changes under different evidential and resource constraints, while the coach-judge condition tests how far externally observed trajectories can be steered during a run.
    \item Across 408 trajectories spanning multiple models, repositories, and launch conditions, we demonstrate that the apparatus makes SWE-agent code-understanding behavior empirically readable through its lenses. The projection lenses expose differences in navigation, synthesis, grounding, and stopping behavior. We also place these observations in dialogue with broader findings on agent trajectories, giving additional support to the proposed method.
\end{enumerate}

The remainder of this paper is organized as follows. Section~\ref{sec:rqs} defines the SWE agent apparatus and formulates the research questions it enables. Section~\ref{sec:methodology} describes the experimental scaffold, including repository-level code-understanding tasks, heterogeneous repository selection, foundation models and launching conditions, and the trajectory recording scheme, followed by the observation lenses through which recorded trajectories are projected into readable form. Section~\ref{sec:results} reports results across three research questions, from surface signals through cross-lens projection to the intervention boundary discovered under coaching. Section~\ref{sec:discussion} situates the findings in the broader literature and discusses what the observation method makes transferable beyond this study. Section~\ref{sec:threats-limits} addresses threats to validity and the epistemic boundaries of the apparatus. Section~\ref{sec:related} surveys related work on trajectory-aware evaluation, tool-mediated interaction, and repository-level code understanding.

\section{Formulation of Research Questions}\label{sec:rqs}

\FloatBarrier

\subsection{The Wild in Bounds: Crafting an Observable Digital World for SWE Agents}\label{sec:wild-bounds}

Before asking how SWE agent mindsets can be projected from behavioral traces, we specify the object and setting that make such trajectories observable.

We use the term SWE agent in a methodological sense. In this paper, an SWE agent is an LLM-mediated, tool-using interaction process. It is activated by a repository-level code-understanding task, acts through a finite sequence of turns in a bounded digital software environment, and terminates when it invokes a distinguished completion tool. We denote a scoped SWE agent as
\begin{equation}\label{eq:swe-agent}
\mathcal{A}_{\theta} = \langle M_{\theta},\; P,\; \mathcal{U},\; u^{*} \rangle,
\end{equation}
where $M_{\theta}$ is the foundation model; $P$ is the prompt and role scaffold that defines the agent's ``framing'' across launch conditions; $\mathcal{U} = \{u_1, \ldots, u_k\}$ is the set of available tools; and $u^{*} \in \mathcal{U}$ is a designated termination tool whose invocation ends the run. Action selection follows the mechanism the foundation model implements given $P$, $\mathcal{U}$, and the accumulated interaction history; we neither assume nor require an explicit, separable policy.

A run is activated by placing this agent in a digital software environment $\mathcal{E}$ under a task $\tau$ and an experimental condition $c$:
\begin{equation}\label{eq:run}
\rho = \mathrm{Run}(\mathcal{A}_{\theta},\; \mathcal{E},\; \tau,\; c).
\end{equation}
The run yields a finite trajectory:
\begin{equation}\label{eq:trajectory}
\xi = \langle\, (x_0, a_0, o_1),\; (x_1, a_1, o_2),\; \ldots,\; x_n \,\rangle, \quad n \leq N,
\end{equation}
where $x_t$ is the accumulated interaction history at turn $t$, $a_t$ is the selected tool call, and $o_{t+1}$ is the observation returned by the environment. The trajectory terminates when $a_t = u^{*}$ or when the turn bound $N$ is reached.

We define the surrounding apparatus as
\begin{equation}\label{eq:apparatus}
\mathcal{D} = \langle\, \mathcal{W},\; \mathcal{T},\; \mathcal{R},\; \mathcal{L} \,\rangle,
\end{equation}
where $\mathcal{W}$ is the repository world, $\mathcal{T}$ is the set of task activators, $\mathcal{R}$ is the recording scheme that preserves each run as an analyzable trajectory, and $\mathcal{L}$ is a set of observation lenses. The repository world $\mathcal{W}$ consists of real software repositories that preserve their directory structure, cross-file references, and architectural layering, with evidence unevenly distributed across these structures. Task activators $\mathcal{T}$ launch the agent with directed but open-ended purposes that require repository-level understanding. The recording scheme $\mathcal{R}$ captures each run as a finite sequence of turns, preserving tool calls, observations, and intermediate reasoning artifacts within each turn.

In this paper, ``\emph{mindset}'' refers to an externally projected stance that is read through the observation lenses. The emerging \emph{mindset} is visible through the trajectories, by looking at them in a proper distance. The lenses create a spectrum of distances for the observers.

Each observation lens $\ell \in \mathcal{L}$ is a mapping:
\begin{equation}\label{eq:lens}
\ell : \Xi \rightarrow \mathcal{V}_\ell
\end{equation}
that projects the trajectory data into a readable view $\mathcal{V}_\ell$, structured for comparison and interpretation, providing a scientific distance, namely a range within which we can observe and compare the agent behaviors under study. In this paper, we demonstrate five lenses: \emph{surface activity view}, \emph{journey arc lens}, \emph{action policy lens}, \emph{automatic adjudication lens}, and \emph{codebase footprint lens}. These range from aggregate statistical summaries to fine-grained projections, both evaluative and spatial, of the trajectory. The set $\mathcal{L}$ is open; the five lenses demonstrated here are a minimal set sufficient to show that trajectory-level observation yields distinguishable behavioral signals. Researchers working with the same apparatus can define additional lenses over the same recorded data.

\begin{figure}
    \centering
    \includegraphics[width=1.0\linewidth]{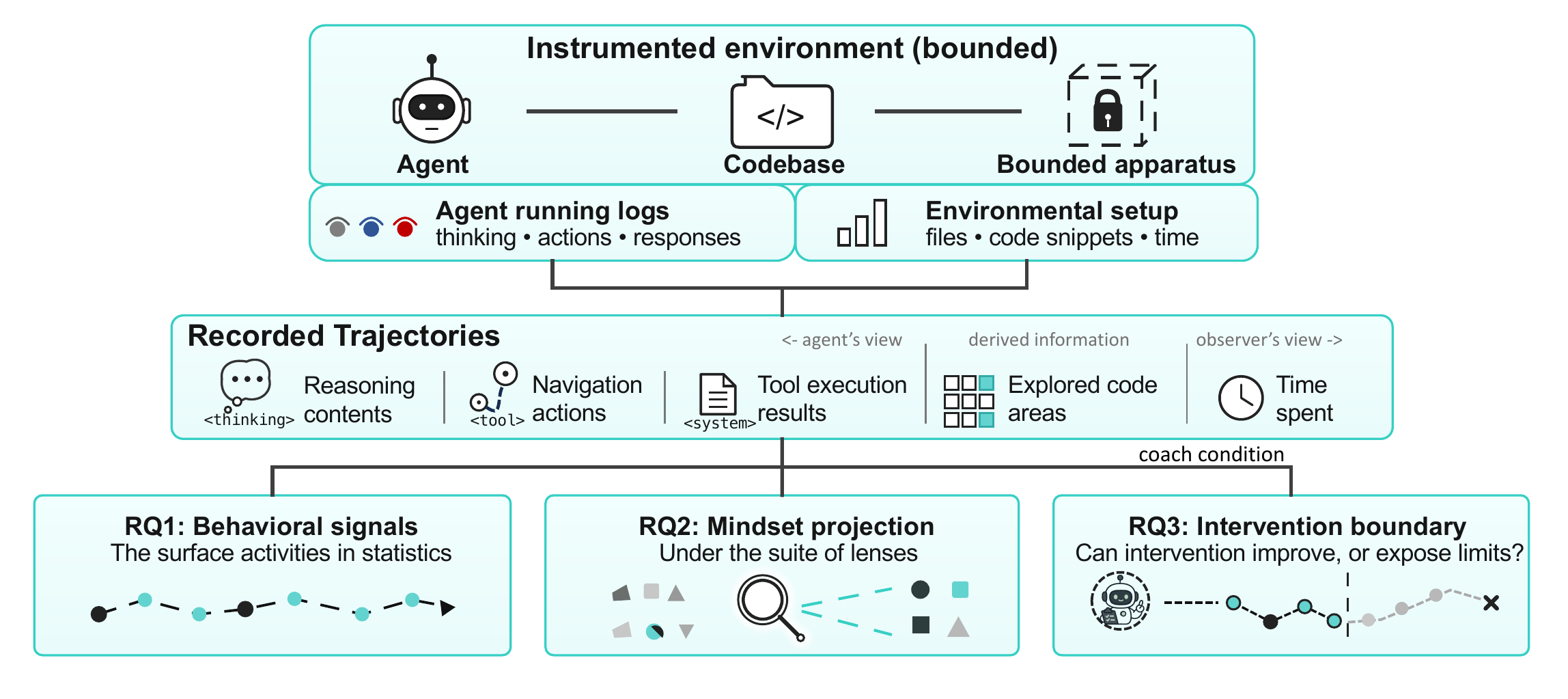}
    \caption{Our bounded instrumented environment records both agent-understanding traces and environment traces, which are recomposed into observed trajectories that ground the paper’s three research questions on behavior, mindset projection, and intervention boundary.}
    \label{fig:bridging-rqs}
\end{figure}

\subsection{Research Questions Enabled by the Apparatus}

The apparatus above does not merely produce task answers. It produces observable trajectories: bounded records of how an SWE agent moves, reads, and reasons inside a repository world, how it stores intermediate understanding along the way, and how it terminates. This shift from final answers to trajectories motivates the research questions of this study.

As illustrated in Figure~\ref{fig:bridging-rqs}, the bounded instrumented environment records both agent-understanding traces and environment traces, which are recomposed into observed trajectories. These trajectories support three connected questions on behavior, mindset projection, and the intervention boundary. First, when an SWE agent is launched into a wild-but-bounded repository world, what behavioral signals are recorded, and what patterns emerge as the launch conditions change? Second, when those signals are interpreted through a mindset observation protocol, what do they reveal beyond aggregate statistical facts, and can we summarize them as behavioral profiles? Third, when an external observer is allowed into the agent channel through guided intervention, does the intervention improve the produced understanding, or does it instead reveal a boundary on what observation and control can perturb at the agentic behavioral level?

\paragraph{RQ1: How do the signals gathered from SWE agent trajectories distribute within our wild-but-bounded environment?} To explore whether the construction of our ``wild digital world'' supports robust observation, we launch \textbf{Ada}, the code-understanding SWE agent, in the experimental apparatus. The prompting status $P$ was initially framed as the \emph{natural} state, with no additional limits applied. And then, we used both \emph{budget pressure} and \emph{prior suppression} to stimulate the agent. Trajectories were recorded. 

\paragraph{RQ2: When projected through mindset observation lenses, what do these behavioral tendencies reveal of the SWE agent and the underlying LLMs?} RQ1 asks whether the apparatus records behavioral signals at all. RQ2 asks whether those signals become meaningful only when read across multiple lenses, particularly in cases where surface statistics fail to explain process-level differences. Using the recorded trajectories from RQ1, we therefore interpret behavior through the ordered lens stack, moving from surface activity to temporal shape, action transitions, semantic adjudication, repository footprint.

If trajectories can be observed from outside, the next question is whether that outside view can improve or steer the ongoing run. We treat this as an empirical question, not an assumption.

\paragraph{RQ3: Under external observer-guided coaching, does the agent improve its answer, or does the intervention expose an intrinsic boundary of mindset?}\label{sec:rq3} RQ2 motivates a third question. Can a trajectory-informed outside intervention steer the run toward better understanding, or does it instead expose a boundary on what external feedback can change? We connect the running agent to an online judging panel, then route panel feedback through a coach who decides when to intervene. In this setup, a run of Ada becomes:

\begin{equation}\label{eq:coach-agent}
    \rho = \mathrm{Run}(\mathcal{A}_{\theta},\; \mathcal{E},\; \tau,\; c_{cj},\; \mathcal{C}).
\end{equation}

The trajectory produced by this run retains the sequential turn structure of Equation~\ref{eq:trajectory}, but the content of each observation $o_{t+1}$ is conditionally augmented. At each turn, a judge panel $\mathcal{J}$ reads the trajectory up to turn $t$ and produces a diagnostic report $\mathcal{J}(\xi_{\leq t})$. The coach $\mathcal{C}$ receives the agent's current action $a_t$ and the panel's report, and selects a decision $d_t \in \{\texttt{PASS},\, \texttt{HINT},\, \texttt{REJECT}\}$:
 
\begin{equation}\label{eq:coach-decision}
d_t = \mathcal{C}\bigl(a_t,\; \mathcal{J}(\xi_{\leq t})\bigr).
\end{equation}
 
The observation is returned to the agent, and subsequent actions depend on this decision.:
 
\begin{equation}\label{eq:coach-observation}
o_{t+1} = 
\begin{cases}
r_{t+1} & \text{if } d_t = \texttt{PASS}, \\
r_{t+1} \oplus m_{\mathcal{C}} & \text{if } d_t \in \{\texttt{HINT},\, \texttt{REJECT}\},
\end{cases}
\end{equation}
 
\noindent where $r_{t+1}$ is the tool execution result, and $m_{\mathcal{C}}$ is the coach's intervention message. Under \texttt{PASS}, the trajectory proceeds as in the main conditions. Under \texttt{HINT}, the coach provides strategic guidance alongside the tool result. Under \texttt{REJECT}, the coach returns the agent's conclusion attempt with a diagnostic message, and the trajectory continues rather than terminating.

\FloatBarrier

\section{Methodology}\label{sec:methodology}

\begin{figure}
    \centering
    \includegraphics[width=1.0\linewidth]{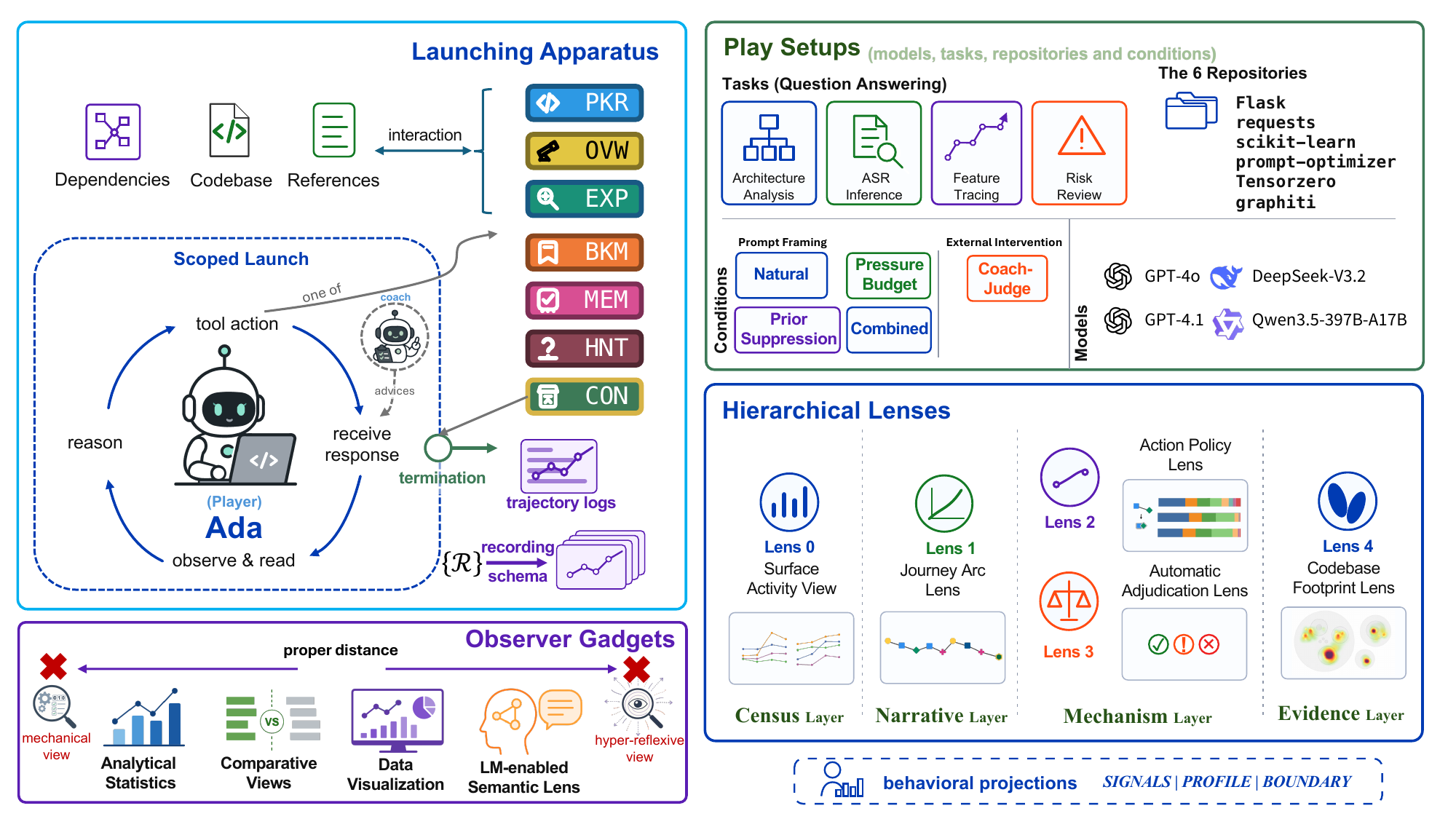}
    \caption{Methodology overview. Within a bounded yet wild repository world, task activators launch Ada under controlled conditions, and the apparatus records its reasoning and tool-use trajectories. Observation lenses then project these trajectories into comparable views, supporting trajectory-level analysis of behavioral signals, mindset projection, and the identification of intervention boundary.}
    \label{fig:method-overview}
\end{figure}

\subsection{Methodological Position: Projecting Code-Understanding Trajectories}

Before describing the experimental scaffold and the analysis protocol, we position the methodological stance that governs both. Repository-level code understanding does not terminate in an executable artifact or a pass-fail verdict. The trajectory of tool-mediated encounters with a repository is the primary empirical object, so the method must be designed around that object.

In this paper, we treat trajectories as traces of observable choice, instead of direct access to internal mindset. The lenses operate as projections from those traces into scoped behavioral profiles.

Trajectories do not interpret themselves. Aggregate statistics describe surface activity, but structured observation through the lenses formally defined in Section~\ref{sec:wild-bounds} is what makes behavioral profiles visible and comparable. Projection, as we use the term, extends beyond computing individual lens outputs. It includes reading across lens results under matched conditions, then anchoring the differences in trajectory-level evidence, before rendering the results visually so that aspects of an emerging code-understanding mindset become available for scientific discussion.

Figure~\ref{fig:method-overview} previews the methodological position we take, including the observational distance at which trajectory reading becomes meaningful. Too close, a mechanical reading of raw tool calls yields activity without behavior, and signals that never resolve into a discussable profile. Too far, an over-interpretive reading turns every trace into motive and loses discriminability. Between these extremes lies a workable range in which interpretations stay accountable to the trace that produced them; we place our observer gadgets in this range, using bounded launches, trace-preserving recording, and lens-based projections to make agent processes observable without treating them as hidden internal state.

The findings reported in this paper should be read as demonstrations that the method works within our bounded experimental environment. The apparatus is designed for structured observation, not competitive ranking. In this sense, Ada is not designed as a high-performance showcase or a benchmarking system for code-reading performance. It is an observation instrument.

By selecting and combining lenses, researchers can build observational instruments suited to viewing intelligent-agent trajectories at different levels of granularity.

\subsection{Launching a Scoped SWE Agent into Code-Understanding Worlds}
 
We scope Ada so that its launches yield trajectories that are comparable across conditions and interpretable through our observation lenses. The scaffold fixes what can be recorded and keeps the account external. Repository access is read-only, so trajectories reflect how the agent reads the world rather than how it modifies it. A small typed tool surface further separates navigation from cognitive bookkeeping, making both observable in the recorded trace.

Ada instantiates the scoped SWE agent $\mathcal{A}_{\theta} = \langle M_{\theta},\; P,\; \mathcal{U},\; u^{*} \rangle$ (Equation~\ref{eq:swe-agent}). We vary $M_{\theta}$ across the models in Section~\ref{sec:experimental-matrix}; the tool set $\mathcal{U}$ is summarized in Table~\ref{tab:ada-tools}; and termination $u^{*}$ is implemented by \texttt{stash\_conclusion}.

The scaffold $P$ follows a ReAct-style \cite{yao2022react} design, informed by publicly available SWE agent prompt architectures \cite{cline2024}. Each turn externalizes a brief pre-action rationale and issues exactly one tool call, yielding a discrete sequence that can be compared across runs. The scaffold also carries forward an accumulating working state (memories, bookmarks, and conclusions), so trajectories show how local evidence is consolidated into intermediate understanding over time. Appendix~\ref{app:experiment-pipeline} details the prompt template and interface.
 
The tool set $\mathcal{U}$ defines Ada's bounded interface to the repository world (Table~\ref{tab:ada-tools}). We choose a small set of typed, developer-like reading actions to make repository contact observable and comparable, while preventing uncontrolled side effects from arbitrary filesystem access. The tools separate navigation (where the agent looks) from cognitive bookkeeping (what it commits as intermediate understanding), a distinction that later lenses use to relate movement through the codebase to stabilization of beliefs.

\begin{table}[t]
\centering
\caption{Ada tool set, grouped by category. Navigation tools interact with the repository; cognitive tools manage the agent's working state; the oracle tool requests researcher clarification; and the termination tool ($u^{*}$) records conclusions and governs session closure.}
\label{tab:ada-tools}
\begin{tabularx}{\linewidth}{%
  p{0.07\linewidth}
  p{0.20\linewidth}
  X
}
\toprule
\textbf{Abbr.} &
\textbf{Tool} &
\textbf{Usage} \\
\midrule
\multicolumn{3}{l}{\textit{Navigation tools}} \\
\addlinespace[0.2em]
OVW &
\texttt{get\_structural\newline \_overview} &
Retrieve the high-level structural outline of a file or a named logical block within it. \\
\addlinespace[0.35em]
PKR &
\texttt{peek\_code\_region} &
Read a specified range of source lines for detailed examination. \\
\addlinespace[0.35em]
EXP &
\texttt{explore\_code} &
Navigate code relationships: jump to a symbol's definition, find all references to a symbol, or advance to the next logical block. \\
\midrule
\multicolumn{3}{l}{\textit{Cognitive tools}} \\
\addlinespace[0.2em]
MEM &
\texttt{create\_memory} &
Record a titled working note synthesizing current understanding, hypotheses, or partial models. \\
\addlinespace[0.35em]
BKM &
\texttt{bookmark} &
Mark a code location with an attached question or reminder for later return. \\
\midrule
\multicolumn{3}{l}{\textit{Oracle tool}} \\
\addlinespace[0.2em]
HNT &
\texttt{get\_hint\_on\newline \_snippet} &
Request a targeted clarification from the researcher on a code snippet that remains unclear after self-directed analysis. \\
\midrule
\multicolumn{3}{l}{\textit{Termination tool} ($u^{*}$)} \\
\addlinespace[0.2em]
CON &
\texttt{stash\_conclusion} &
Record a formal conclusion about the code and signal whether to continue analysis or finalize the session. \\
\bottomrule
\end{tabularx}
\end{table}

Termination is designed as part of the observation record. A run ends when the agent submits a concluding account via \texttt{stash\_conclusion} (the distinguished tool $u^{*}$ in Equation~\ref{eq:swe-agent}), or when the predefined turn-bound $N$ is reached; separating intermediate synthesis from explicit closure keeps both analyzable in the trajectory.
 
Before scaling the experiment, three prompt-scaffold variants were tested under the same tool design. All produced acceptable code-understanding accounts and reached explicit termination, confirming that the apparatus supports finite, recordable trajectories. The stricter ($N = 100$) turn bound was applied later in the main matrix. Detailed pilot materials are reported in Appendix~\ref{app:pilot-materials}.

\subsection{Experimental Matrix: Tasks, Repositories, Models, and Conditions}\label{sec:experimental-matrix}

Repository-level code understanding unfolds through trajectories, and trajectories become scientifically comparable only when the launch conditions that produce them are specified across fixed axes. The experimental matrix varies over four axes over the constant scaffold described in the previous subsection: task family, repository, foundation model, and experimental condition. This structure makes the resulting trajectories part of a single observation design rather than isolated demonstrations.

Behavioral signals become visible only when runs can be aligned along shared launch axes. Tasks provide the most natural separators of initial ``kick-offs'' on an agent's journey, while prompt-framing conditions introduce controlled constraints that scope how the foundation model justifies claims and decides to commit to them. The coaching condition then moves from agent-internal setups to intervention by admitting an external observer into the agent channel and testing the boundary between what can be read from traces and what can be steered. Taken together, the matrix is deliberately scoped. It covers factors that SWE agent designers plausibly care about while avoiding implementation-level variations.

The task family axis $\mathcal{T}$ specifies the initiating forces that bring the agent into a code-understanding situation, while leaving the ensuing route of exploration under agent control. We draw four task families from recognizable software-engineering situations (Table~\ref{tab:task-families}) \cite{swebok2014} so that trajectories can be compared across distinct kinds of understanding work rather than a single prompt template.

The repository axis $\mathcal{W}$ specifies the codebase worlds into which the agent is launched (Table~\ref{tab:repository-fields}). We intentionally mix system types, scales, and organizational styles so that observed behaviors are not artifacts of one architectural idiom, but stable enough to support cross-repository comparison.

\begin{table}[t]
\centering
\caption{Task families and the software-engineering situations they reflect. Each family targets a distinct form of repository-level code understanding, grounded in a recognizable moment in real software work.}
\label{tab:task-families}
\begin{tabularx}{\linewidth}{p{0.22\linewidth} p{0.28\linewidth} X}
\toprule
\textbf{Task Family} & \textbf{Software-Engineering Situation} & \textbf{Code-Understanding Challenge} \\
\midrule
\emph{architecture analysis} &
Onboarding, dependency evaluation, migration planning &
Identify major components, architectural style, and interaction patterns across the repository. \\
\addlinespace[0.35em]
\emph{ASR inference} &
Architectural reverse-engineering in underdocumented or legacy codebases &
Reconstruct implicit architectural requirements from code evidence. \\
\addlinespace[0.35em]
\emph{feature implementation tracing} &
Bug localization, change-impact analysis, feature-level debugging &
Locate and describe the end-to-end implementation path of a specified feature. \\
\addlinespace[0.35em]
\emph{risk and code-smell review} &
Code review, maintenance triage, technical-debt assessment &
Identify risks, comprehension difficulties, or code smells with supporting code locations. \\
\bottomrule
\end{tabularx}
\end{table}

\begin{table}[t]
\centering
\caption{Repository fields used in the Ada apparatus.}
\label{tab:repository-fields}

\begin{tabularx}{\linewidth}{l l r r l}
\toprule
\textbf{Repository} &
\textbf{Language} &
\textbf{Files} &
\textbf{LOC} &
\textbf{System Type} \\
\midrule

Requests & Python & 29 & 8{,}060 & HTTP client library \\
\multicolumn{5}{p{\linewidth}}{\emph{Role:} Tests implementation tracing behind a focused public API.} \\
\addlinespace[0.6em]

Flask & Python & 29 & 9{,}526 & Web framework \\
\multicolumn{5}{p{\linewidth}}{\emph{Role:} Tests framework abstraction, context mechanisms, routing, and extension structure.} \\
\addlinespace[0.6em]

Scikit-learn & Python & 554 & 258{,}342 & Scientific ML library \\
\multicolumn{5}{p{\linewidth}}{\emph{Role:} Tests structural compression across a broad algorithmic ecosystem.} \\
\addlinespace[0.6em]

Graphiti & Python & 109 & 14{,}321 & Agent-memory infrastructure \\
\multicolumn{5}{p{\linewidth}}{\emph{Role:} Tests graph, LLM, retrieval, and service coordination.} \\
\addlinespace[0.6em]

Prompt Optimizer & TypeScript & 172 & 30{,}694 & Prompt-engineering product platform \\
\multicolumn{5}{p{\linewidth}}{\emph{Role:} Tests frontend, service layer, template, storage, and workflow integration.} \\
\addlinespace[0.6em]

TensorZero & Rust & 438 & 149{,}755 & LLMOps platform \\
\multicolumn{5}{p{\linewidth}}{\emph{Role:} Tests provider abstraction, inference, observability, evaluation, and optimization loops.} \\

\bottomrule
\end{tabularx}
\end{table}

Varying the foundation model $M_\theta$ provides a comparison axis while holding the scaffold fixed, making differences in evidence use, exploration scope, and conclusion grounding observable in trajectories.

The main experimental matrix instantiates Ada with four foundation models (GPT-4o\footnote{OpenAI model documentation: \url{https://developers.openai.com/api/docs/models/gpt-4o}.}, GPT-4.1\footnote{OpenAI announcement: \url{https://developers.openai.com/api/docs/models/gpt-4.1}.}, DeepSeek-V3.2\footnote{Model card (PDF): \url{https://fe-static.deepseek.com/chat/transparency/deepseek-v3.2-model-card-0414-EN.pdf}.}, and Qwen3.5-397B-A17B\footnote{Model card: \url{https://huggingface.co/Qwen/Qwen3.5-397B-A17B}.}), chosen to provide bounded diversity rather than exhaustive coverage. Models that could not reliably sustain the tool-call protocol in pilot runs were excluded from trajectory generation.\footnote{The GLM family was excluded from trajectory generation due to unreliable tool-call compliance, and was used only in the post-hoc evaluation layer. See Appendix~\ref{app:pilot-materials}.}

For readability, we later refer to DeepSeek-V3.2 as \emph{DeepSeek} and Qwen3.5-397B-A17B as \emph{Qwen}, unless otherwise specified. GPT-4.1 and GPT-4o are kept as-is.

The condition axis $c$ varies prompt-framing conditions over the same scaffold, so cross-condition comparisons isolate changes in guidance and evidential stance rather than changes in tools or repository access. \emph{Budget pressure} operationalizes resource-aware launching to test how constrained planning reshapes trajectory organization \cite{simon1955behavioral,shinn2023reflexion}. \emph{Prior suppression} enforces an evidence-first stance to separate trace-grounded understanding from prior familiarity, without attempting to remove knowledge from the model in the sense of unlearning \cite{bourtoule2021unlearning}. Finally, the coach-judge condition introduces an external observer into the agent channel to probe the boundary between what can be read from trajectories and what can be steered. Table~\ref{tab:experiment-conditions} summarizes the conditions and their observation roles (the core grid yields \(\lvert\mathcal{W}\rvert \times \lvert\mathcal{T}\rvert \times \lvert M_{\theta}\rvert \times \lvert c\rvert = 6 \times 4 \times 4 \times 4 = 384\) trajectories).

\begin{table}[t]
\centering
\caption{Experimental conditions. The core grid (top) systematically varies budget and prior-knowledge constraints across all four models and six repositories. The intervention condition (bottom) introduces an external coaching mechanism and is used exclusively for RQ3.}
\label{tab:experiment-conditions}

\begin{tabularx}{\linewidth}{%
  p{0.20\linewidth}
  p{0.34\linewidth}
  X
}
\toprule
\textbf{Condition} &
\textbf{What's Different} &
\textbf{Observation Purpose} \\
\midrule
\multicolumn{3}{l}{\textit{Prompt-framing conditions (RQ1, RQ2)}} \\
\addlinespace[0.2em]

\emph{natural condition} &
Standard launch; no special settings applied &
Provides the unperturbed reference trajectory against which all other conditions are compared. \\
\addlinespace[0.35em]

\emph{budget pressure} &
Explicit 12-turn budget awareness with progressive convergence reminders &
Tests whether trajectory organization restructures when exploration space is compressed. \\
\addlinespace[0.35em]

\emph{prior suppression} &
Evidence-first constraint treating prior familiarity as weak intuition rather than sufficient ground &
Makes the boundary between prior-driven and trace-driven reasoning more observable. \\
\addlinespace[0.35em]

\emph{combined condition} &
\emph{Budget pressure} and \emph{prior suppression} applied together &
Provides an interaction reference for observing both constraints in the same trajectory. \\

\midrule
\multicolumn{3}{l}{\textit{External intervention condition (RQ3)} {\footnotesize (only for GPT-4.1)}} \\
\addlinespace[0.2em]

\emph{coach-judge} &
An external coaching and judging mechanism intervenes during the agent's reasoning process &
Tests whether external intervention improves or degrades what the agent's mindset produces. \\

\bottomrule
\end{tabularx}
\end{table}

The four prompt-framing conditions in the core grid are summarized in Table~\ref{tab:experiment-conditions} and provide the main comparison set for RQ1 and RQ2.

The four main conditions keep each trajectory self-directed. The coach-judge condition crosses that boundary by introducing an external coach whose message can enter the agent's observation stream, based on a judge panel reading of the trajectory so far. The intervention mechanism is defined in Equation~\ref{eq:coach-agent} (Section~\ref{sec:rq3}), and the pipeline details are given in Appendix~\ref{app:coach-judge-pipeline}.

The experimental matrix, therefore, defines the apparatus's bounded empirical range. The 384 main-matrix trajectories and 24 coach-judge trajectories together constitute the 408-trajectory analysis object from which the observation lenses, introduced next, construct behavioral profiles.

\subsection{Analysis Lenses: A Way of Seeing}\label{sec:lenses}

Figure~\ref{fig:method-overview} summarizes the analysis stack that connects raw trajectories to the claims reported in this paper. The \emph{observer gadgets} provide the intermediate instrumentation: they operationalize aspects of agent behavior and transform long, tool-mediated traces into stable, comparable \emph{views} that support systematic reading across models, prompting conditions, repositories, and task families. Concretely, an observation lens $\ell \in \mathcal{L}$ is defined as a map from trajectory evidence to a structured view, $\ell : \Xi \rightarrow \mathcal{V}_\ell$ (Section~\ref{sec:wild-bounds}). The lenses are designed for empirical analysis, i.e., for researchers who must compare and aggregate behaviors across many runs rather than interpret individual trajectories in isolation. We instantiate five lenses as a working set; together they suffice to demonstrate that trajectory-level observation yields distinguishable behavioral profiles across models and conditions.

Although individual trajectories can be interpreted through close reading, this approach does not scale to our 408-run corpus and does not yield observations that are comparable across models and experimental conditions. We therefore record each run using a standardized scheme $\mathcal{R}$ that preserves the turn structure and artifacts required for later projection, then treat the resulting traces as empirical material that becomes evidence only when analyzed through lenses.

The lenses are ordered because different observations carry different levels of interpretive risk. Surface activity stays closest to the trace. Temporal shape, action transitions, semantic judgment, and repository footprint require progressively stronger interpretation.

Because no single perspective is adequate for characterizing agent behavior, we organize the lenses hierarchically into four layers (Figure~\ref{fig:method-overview}): the \emph{census layer}, the \emph{narrative layer}, the \emph{mechanism layer}, and the \emph{evidence layer}. The purpose of this layering is to support progressive reading: corpus-level summaries enable screening and comparison, while increasingly specific projections recover temporal structure, decision mechanisms, and repository-grounded evidence needed to substantiate explanatory claims.

In the census layer, lens~0 (the \emph{surface activity view}) summarizes each trajectory into activity statistics that establish a baseline for comparing models and conditions across the run matrix. It separates changes in overall effort from changes in higher-level behavior. Its summary values recur as reference points for the other lenses.

In the narrative layer, lens~1 (the \emph{journey arc lens}) renders each trajectory on a normalized timeline so that many runs can be compared in a common visual frame. By foregrounding temporal structure instead of isolated events, it supports corpus-scale interpretation through recurrent phases, loops, and atypical trajectories.

In the mechanism layer, lens~2 (the \emph{action policy lens}) characterizes tool use as sequential decision behavior by modeling adjacent tool transitions as $n$-gram patterns \cite{bouzenia2025trajectories}. Table~\ref{tab:transition-codes} defines the transition codes used in this work; the coding is apparatus-specific because it depends on Ada's tool vocabulary and on the meaning of tool transitions in a code-understanding setting.

\begin{table*}[t]
\centering
\caption{Action policy codes and readings.}
\label{tab:transition-codes}

\small
\begin{tabularx}{\textwidth}{%
  p{0.17\textwidth}
  p{0.30\textwidth}
  X
}
\toprule
\textbf{Action Policy} &
\textbf{Operational Rule(s)} &
\textbf{Interpretive Reading} \\
\midrule

close read 
&
\texttt{PKR $\rightarrow$ PKR\$; EXP $\rightarrow$ PKR\$} &
Sustained inspection of adjacent or continuous code regions. \\
\addlinespace[0.3em]

skim 
&
\texttt{OVW $\rightarrow$ OVW; NAV $\rightarrow$ PKR*; ANY $\rightarrow$ PKR*} &
Broad sampling across structures or file boundaries. \\
\addlinespace[0.3em]

traverse 
&
\texttt{NAV $\rightarrow$ PKR; ANY $\rightarrow$ PKR} &
Ordinary movement from navigation into code reading. \\
\addlinespace[0.3em]

trace 
&
\texttt{ANY $\rightarrow$ EXP; ANY $\rightarrow$ OVW*} &
Following code relations or shifting structural attention across files. \\
\addlinespace[0.3em]

zoom in 
&
\texttt{OVW $\rightarrow$ PKR; OVW $\rightarrow$ EXP} &
Moving from structural overview toward local inspection or traversal. \\
\addlinespace[0.3em]

zoom out 
&
\texttt{ANY $\rightarrow$ OVW} &
Returning from local action to structural overview. \\
\addlinespace[0.3em]

resume 
&
\texttt{COG $\rightarrow$ NAV} &
Returning from cognitive work to repository exploration. \\
\addlinespace[0.3em]

other read 
&
\texttt{ANY $\rightarrow$ NAV} &
Residual navigation transition not captured by more specific rules. \\
\addlinespace[0.3em]

close out 
&
\texttt{NAV $\rightarrow$ CON terminal} &
Moving from repository navigation into final synthesis. \\
\addlinespace[0.3em]

interim synth 
&
\texttt{NAV $\rightarrow$ CON non-terminal} &
Recording a provisional conclusion without terminating. \\
\addlinespace[0.3em]

dump 
&
\texttt{NAV $\rightarrow$ COG; ANY $\rightarrow$ COG} &
Externalizing notes, bookmarks, hints, or conclusions after prior action. \\
\addlinespace[0.3em]

cognitive streak 
&
\texttt{COG $\rightarrow$ COG} &
Consecutive consolidation, deferral, clarification, or synthesis actions. \\
\addlinespace[0.3em]

other 
&
\texttt{ANY $\rightarrow$ ANY} &
Fallback for transitions not matched by higher-precedence rules. \\

\bottomrule
\end{tabularx}

\vspace{0.4em}
\footnotesize
\emph{Note:} \texttt{NAV = \{OVW, PKR, EXP\}} and \texttt{COG = \{MEM, BKM, HNT, CON\}}.
A \texttt{*} condition requires different current file targets.
A \texttt{\$} condition requires continuous line ranges.
Rules are applied by precedence: exact rules before wildcard rules; more specific wildcard classes before less specific ones; \texttt{\$} before \texttt{*}; \texttt{*} before regular wildcard rules; remaining ties by list order.

\end{table*}

In the mechanism layer, lens~3 (the \emph{automatic adjudication lens}) adds semantic evaluation that cannot be recovered reliably from structural signals alone. Using LLM-based judging, it assesses whether a trajectory's outcome is relevant to the task, grounded in repository evidence, appropriately calibrated at termination, and navigationally well-oriented; details and validation are reported in Appendix~\ref{app:llm-eval-align}.

In the evidence layer, lens~4 (the \emph{codebase footprint lens}) makes repository contact explicit by projecting each trajectory into a map of visited files and code regions (and their aggregate heatmaps). This view distinguishes focused evidence acquisition from diffuse or shallow exploration, grounding behavioral claims in where the agent actually interacted with the codebase.

The lenses are intended to be read in combination rather than as stand-alone diagnostics. Lens~0 establishes a baseline of effort and tool use; lenses~1--2 recover temporal and sequential structure; lens~3 evaluates conclusion-level adequacy; and lens~4 grounds these readings in repository contact. Agreement across lenses strengthens interpretation; disagreements localize where a behavioral account is underdetermined, motivating closer inspection. The behavioral profiles in Section~\ref{sec:results} are derived from this triangulation, and the set $\mathcal{L}$ remains extensible for additional questions and agent settings.

\section{Results}\label{sec:results}

The results are organized to mirror the layered lens stack introduced in Section~\ref{sec:lenses}. We begin with the \emph{census layer} (RQ1, Section~\ref{sec:res-rq1}), reporting surface signals that establish a baseline of effort and exposure. We then project trajectories through the narrative, mechanism, and evidence layers (RQ2, Section~\ref{sec:res-rq2}), moving progressively from temporal shape through action transitions and semantic adjudication to repository footprints. Finally, RQ3 (Section~\ref{sec:res-rq3}) tests whether the intervention improves outcomes or exposes a boundary of the apparatus.

Throughout, we report observer-oriented findings in a consistent three-part form: \textsc{signals} (what the lens makes measurable), \textsc{profile} (the comparative pattern the projection reveals), and \textsc{boundary} (where the projection ceases to support a stable reading).

\subsection{RQ1: Surface Signals from the Apparatus}\label{sec:res-rq1}

\begin{figure}[H]
\centering
\includegraphics[width=1.0\textwidth]{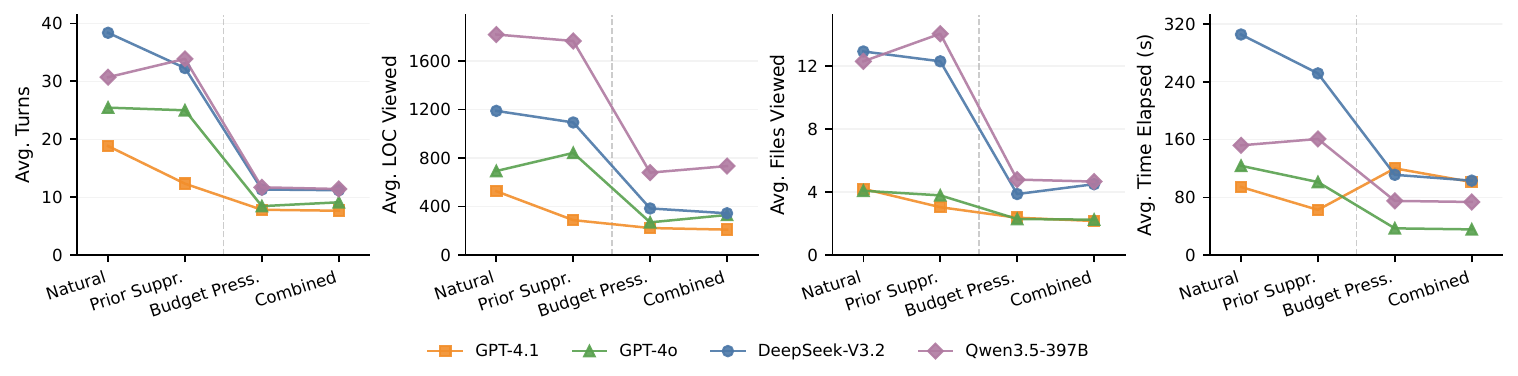}
\caption{Surface-level exploration activity declines markedly under budget pressure and combined constraints across all four LLM backends. Each panel tracks averages.\ turns, LOC viewed, files viewed, and time elapsed (left to right) across four constraint conditions: natural, prior suppression, budget pressure, and combined.}
\label{fig:rq1-surface}
\end{figure}

\begin{figure}[H]
\centering
\includegraphics[width=0.5\textwidth]{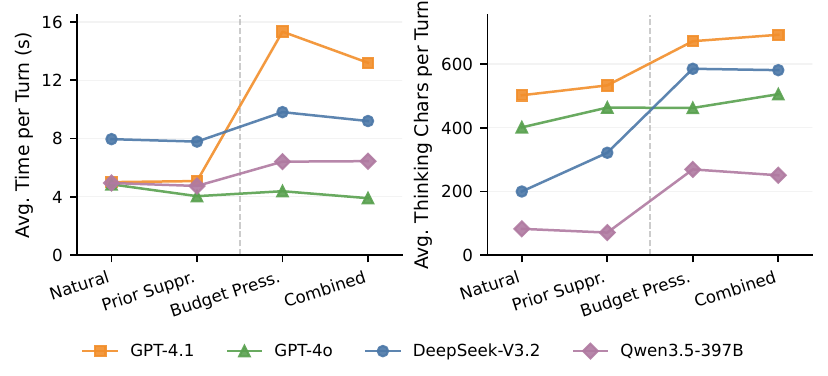}
\caption{Per-turn working effort intensifies under budget pressure, revealing a trade-off between session breadth and turn-level depth. Two panels show the average time per turn (s) and the average thinking characters per turn across natural, prior suppression, budget pressure, and combined conditions for four models.}
\label{fig:rq1-effort}
\end{figure}

Lens~0 (the surface activity view) projects each trajectory into a small set of census-layer signals, namely turns, lines of code viewed, files visited, elapsed time, and per-turn effort, to support cross-model and cross-condition comparison (Figures~\ref{fig:rq1-surface}--\ref{fig:rq1-effort}). At this level, models differ substantially in exploration scale under the natural condition, while budget pressure produces a consistent compression of surface activity across the run matrix.

Figure~\ref{fig:rq1-surface} makes the condition effects explicit. Under budget pressure, turns, LOC viewed, files visited, and total elapsed time decrease for all four models. Under the combined condition, models converge to similarly low activity levels, reducing the between-model spread visible in the natural condition. By contrast, the prior suppression condition yields surface profiles that remain close to the natural baseline across all four panels, indicating that this constraint is not distinguishable at the census layer.

Budget pressure does not merely reduce the number of turns; it also changes per-turn behavior. Figure~\ref{fig:rq1-effort} shows an inversion in the census-layer signals: aggregate activity decreases, while per-turn time and per-turn ``thinking'' characters increase. The increase is most pronounced for GPT-4.1 (time per turn rising from about 5 seconds to over 15 seconds), and is also visible for DeepSeek and Qwen, while GPT-4o remains comparatively stable. Lens~0 makes this trade-off observable, but the trajectory record at this layer does not permit attributing the additional per-turn time to any specific source (e.g., model deliberation versus serving latency).

\begin{finding}
\textbf{\textsc{signals}:} Lens~0 projects each run into surface activity and effort measures (trajectory length, repository exposure, elapsed time, and per-turn effort).
\textbf{\textsc{profile}:} Under the natural condition, models separate by exploration scale, whereas budget pressure produces a consistent compression of surface activity and a concurrent increase in per-turn effort.
\textbf{\textsc{boundary}:} Prior suppression is not distinguishable at this layer, and surface signals do not determine whether observed time increases reflect agent deliberation or external latency.
\end{finding}

\subsection{RQ2: Projecting Trajectories through Observation Lenses}\label{sec:res-rq2}

Beyond the census-layer surface signals in RQ1, RQ2 projects trajectories through the narrative, mechanism, and evidence layers to make observer-readable structure explicit. Across these projections, models and constraints can exhibit markedly different temporal organization, transition grammars, semantic adequacy, and repository contact patterns, even when surface activity appears similar. We report these results in four views: journey arcs (Section~\ref{sec:res-rq2-lens-1}), action policy transitions (Section~\ref{sec:res-rq2-lens-2}), automatic adjudication (Section~\ref{sec:res-rq2-lens-3}), and codebase footprints (Section~\ref{sec:res-rq2-lens-4}).

\subsubsection{Journey Arc: Temporal Shape and Within-Condition Variance}\label{sec:res-rq2-lens-1}
\begin{figure}[t]
  \centering
  
  \begin{subfigure}[t]{0.5\linewidth}
    \centering
    \includegraphics[width=\linewidth]{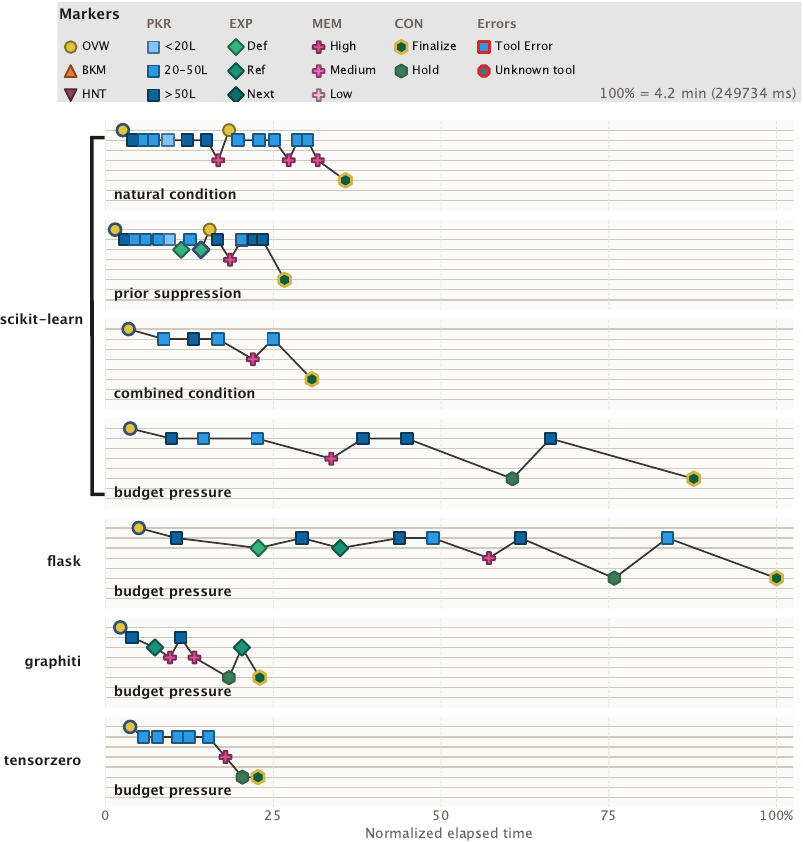}
    \captionsetup{width=0.8\linewidth}
    \caption{Journey arcs of GPT-4.1, under the task of risk and code-smell review, varying in conditions (arc 1--4) and repositories (arc 4--7).}
    \label{fig:rq2-arc-a}
  \end{subfigure}%
  \begin{subfigure}[t]{0.5\linewidth}
    \centering
    \includegraphics[width=\linewidth]{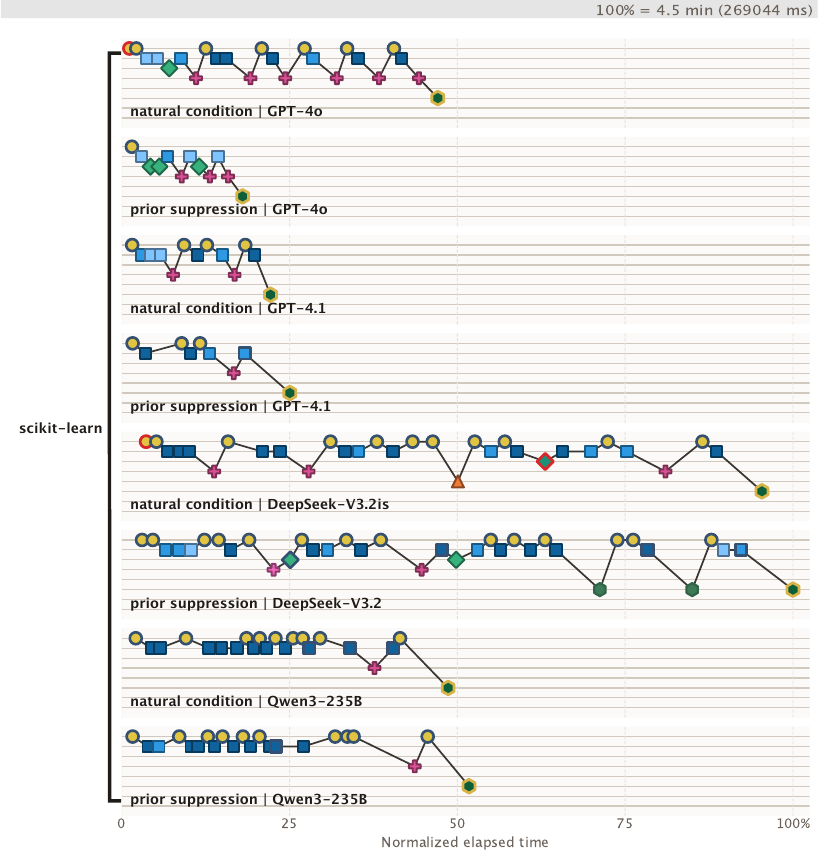}
    \captionsetup{width=0.8\linewidth}
    \caption{Journey arcs under the task of architecture analysis, varying all 4 models, pairing the natural and prior-suppression conditions.}
    \label{fig:rq2-arc-b}
  \end{subfigure}

  \caption{Journey arc lens examples. Each arc is a readable timeline with tool calls shown as markers, elapsed time running left to right, and the full journey visible as an overall contour.}
  \label{fig:res-lens-1}
\end{figure}

The journey arc lens (lens~1) projects each trajectory into a normalized, observer-readable timeline, making temporal organization and within-condition variance directly comparable across runs. Figure~\ref{fig:res-lens-1} shows that constraints can change not only how much activity occurs, but also how that activity is distributed over time.

Figure~\ref{fig:rq2-arc-a} traces GPT-4.1 trajectories for \emph{risk and code-smell review} across conditions. Under budget pressure and the combined constraint, arcs are visibly shorter than under the natural condition, consistent with the census-layer compression in RQ1. In addition, the temporal projection reveals heterogeneous within-trajectory pacing: some budget-pressure runs contain extended gaps between successive actions without a commensurate increase in visible output.

Figure~\ref{fig:rq2-arc-b} contrasts natural and prior suppression across models for \emph{architecture analysis}. Within each model, the two conditions exhibit closely matched arc shapes, whereas the dominant differences are model-specific temporal signatures (e.g., more extended orientation-heavy arcs versus compact, direct arcs).

The temporal gaps visible in some arcs raise an attribution boundary for observer-facing interpretation. For GPT-4.1, per-turn elapsed wall-clock time correlates only weakly with per-turn output length (Pearson $r = 0.397$; Spearman $\rho = 0.532$), so extended pauses in Figure~\ref{fig:rq2-arc-a} cannot be read unambiguously as additional agent work. We therefore treat unusually long per-turn intervals as a limitation of the projection under API-mediated execution; possible serving-stack contributors are discussed in Section~\ref{sec:dis-align-works}.
 
Figure~\ref{fig:rq2-arc-b} pairs natural condition and prior suppression trajectories for all four models on \emph{architecture analysis} in scikit-learn. Within each model, the two conditions produce very similar journey shapes, making them hard to tell apart visually. The main contrast is model-level. GPT-4o and GPT-4.1 remain compact, DeepSeek spans the timeline with OVW-heavy orientation, and Qwen falls in between. At this level of reading, prior suppression leaves Ada's outward journey largely unchanged. In other words, the condition appears to alter evidential stance without reshaping the visible arc of exploration.

\begin{finding}
\textbf{\textsc{signals}:} Lens~1 projects each trajectory as a time-ordered arc, preserving within-run sequencing and per-turn timing.
\textbf{\textsc{profile}:} The projection makes within-condition variance visible. Budget pressure yields shorter arcs and, in some runs, uneven pacing with long inter-action gaps, while natural and prior suppression arcs within a model are visually similar relative to the larger model-to-model differences.
\textbf{\textsc{boundary}:} Wall-clock timing in the arcs is not uniquely attributable to agent work; extended pauses should be treated as an execution artifact boundary under API-mediated serving.
\end{finding}

\subsubsection{Action Policy: Transition Signatures across Models}\label{sec:res-rq2-lens-2}
\begin{figure} 
    \centering
    \includegraphics[width=0.6\linewidth]{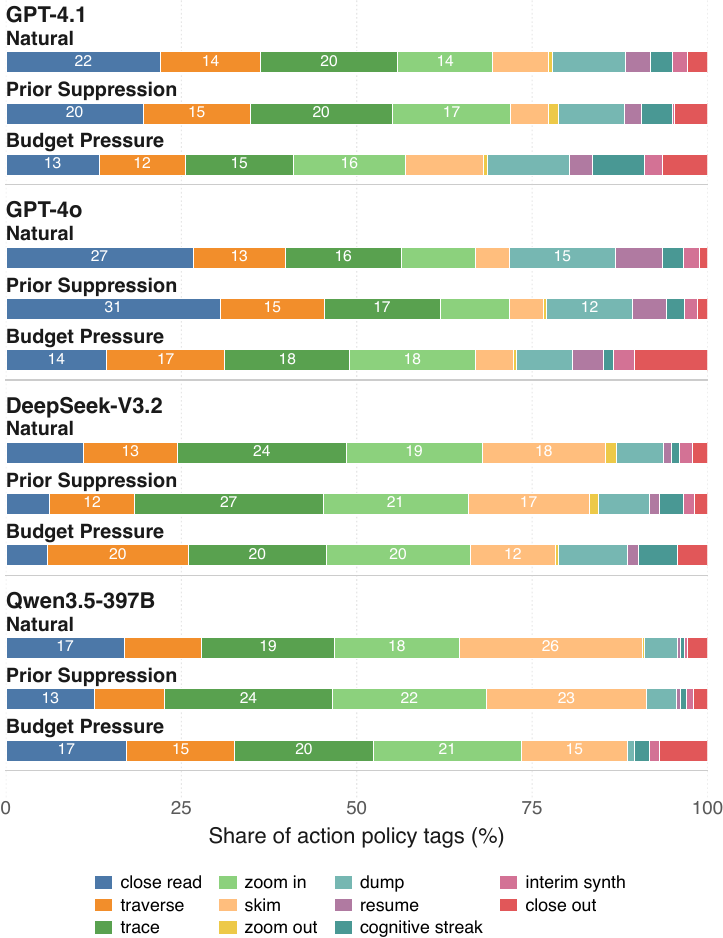}
    \caption{Action policy lens: transition-tag shares across models and conditions. Each row summarizes one model under natural, prior suppression, and budget pressure. Stacked bars show the percentage composition of adjacent tool-to-tool transitions, grouped into the transition tags defined in Table~\ref{tab:transition-codes} (e.g., close read, traverse, skim, zoom, dump, cognitive streak, close out, and intent switch).}
    \label{fig:rq2-policy}
\end{figure}

Across all 384 trajectories, raw tool-call counts are highly concentrated. \texttt{peek\_code\_region} and \texttt{get\_structural\_overview} together account for 80.1\% of all tool calls, and this concentration holds across models and conditions (Appendix~\ref{app:unreported-details}). Lens~2 (the action policy lens), therefore, shifts the reading from marginal tool frequencies to sequential structure by coding adjacent tool-to-tool transitions (Table~\ref{tab:transition-codes}). Figure~\ref{fig:rq2-policy} shows that these transition profiles are model-characteristic, and that budget pressure reshapes them in distinct, model-specific ways rather than producing a single uniform ``compressed'' policy.

Figure~\ref{fig:rq2-policy} reports transition-tag shares by model and condition. It supports two observer-facing readings: models exhibit stable transition signatures, and constraints reshape these signatures in model-specific ways.\footnote{The combined condition is omitted from this figure for compactness. Its profiles closely resemble budget pressure across all models and are reported in Appendix~\ref{app:unreported-details}.}

\emph{Model-wise signatures.} In Figure~\ref{fig:rq2-policy}, GPT-4o emphasizes close reading; DeepSeek is traverse-heavy in line with its OVW-style orientation sequences; Qwen exhibits the highest skimming share; and GPT-4.1 is comparatively evenly distributed across tags.

\emph{Condition-wise shifts.} Natural and prior suppression remain closely matched within each model. Budget pressure changes the transition mix, but without a single common direction: GPT-4.1 shifts toward cognitive transitions and away from close reading; GPT-4o reduces close reading while increasing dump and zoom-in; DeepSeek becomes more evenly distributed; and Qwen trades skim for zoom-in.\footnote{Close-out share rises mechanically under budget pressure because shorter trajectories contain the same single termination action against fewer total transitions. This is a denominator effect, not a behavioral shift.}

Task family also modulates these transition profiles in consistent directions, and the task-stratified results are reported in Appendix~\ref{app:unreported-details}.

\begin{finding}
\textbf{\textsc{signals}:} Lens~2 projects trajectories into transition-tag shares over adjacent tool-to-tool moves (Table~\ref{tab:transition-codes}; Figure~\ref{fig:rq2-policy}).
\textbf{\textsc{profile}:} The resulting transition signatures are model-characteristic, vary systematically with task family, and remain similar between natural and prior suppression. Budget pressure changes the transition mix, but the direction of change depends on the model rather than following a single uniform compression pattern.
\textbf{\textsc{boundary}:} Transition profiles describe sequential structure but do not evaluate whether a given action policy produces higher-quality conclusions; this requires semantic adjudication in lens~3.
\end{finding}

Models navigate differently and reorganize differently under pressure. Whether these different action policies produce different quality outcomes is a question the action policy lens cannot answer. This is the opening for the adjudication lens.

\subsubsection{Adjudication: Conclusion Quality and Epistemic Grounding}\label{sec:res-rq2-lens-3}

\begin{figure}[t]
  \centering
  \includegraphics[width=0.7\linewidth]{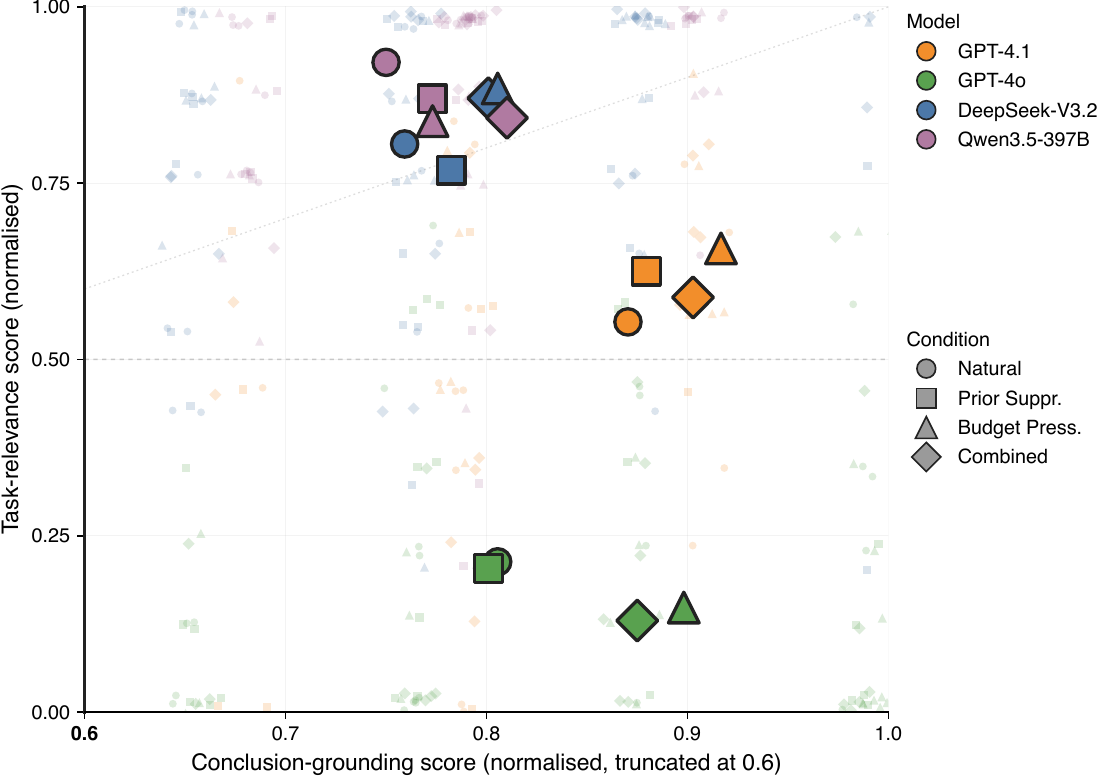}
  \caption{Automatic adjudication scores for \emph{task relevance} and \emph{conclusion grounding}, normalized to $[0,1]$. Points are aggregated by model and condition, showing how conclusion quality varies across the run matrix. Color encodes the model, and marker encodes the condition.}
  \label{fig:rq2-scatter}
\end{figure}

\begin{figure}[t]
  \centering
  \includegraphics[width=\linewidth]{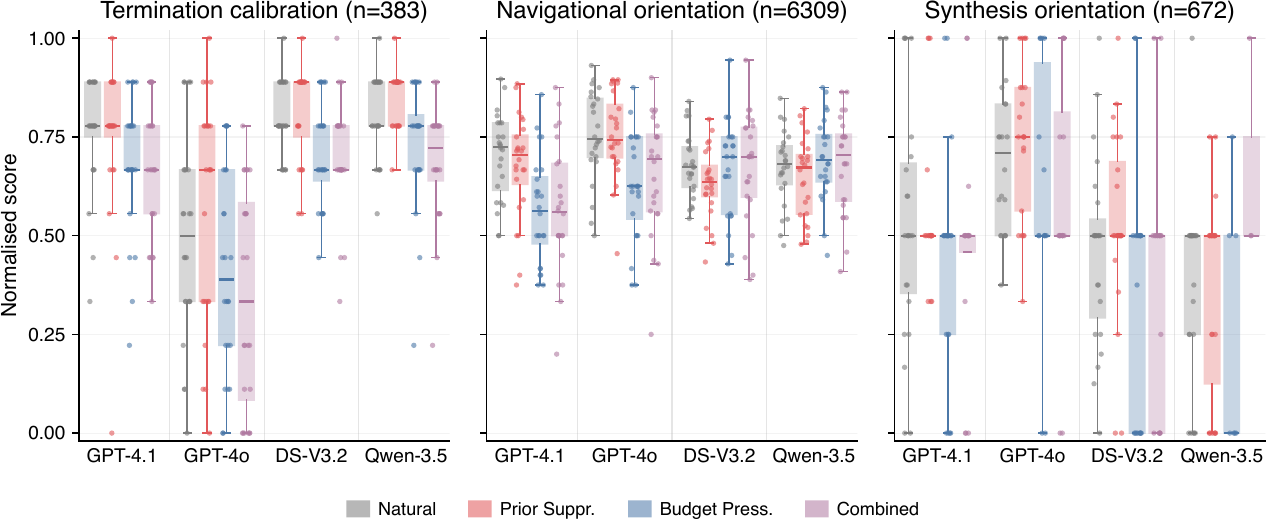}
  \caption{Automatic adjudication scores for \emph{termination calibration}, \emph{navigational orientation}, and \emph{synthesis orientation}, shown left to right. Boxplots summarize per-trajectory distributions by model and condition, and points show individual trajectories. The figure highlights how these dimensions shift under different constraints.}
  \label{fig:rq2-adj-boxplot}
\end{figure}

To evaluate whether different projected policies yield different outcomes, lens~3 (adjudication) applies LLM-based judges to each trajectory conclusion on five dimensions: \emph{task relevance}, \emph{conclusion grounding}, \emph{termination calibration}, \emph{navigational orientation}, and \emph{synthesis orientation}. Figures~\ref{fig:rq2-scatter}--\ref{fig:rq2-adj-boxplot} show that adjudicated quality is primarily model-characteristic, while constraints produce more localized shifts (most clearly in termination calibration under pressure).

Figure~\ref{fig:rq2-scatter} summarizes conclusion-level quality by plotting \emph{task relevance} against \emph{conclusion grounding} for each model-condition pair. The primary pattern is model-wise clustering: T1 ranges from 0.174 (GPT-4o) to 0.868 (Qwen), while condition-level variation remains within 0.608--0.632. A closer reading highlights GPT-4o as an outlier: despite high trace grounding (T2\,=\,0.845) and strong close-reading share in lens~2, its task relevance is substantially lower than the other models.

A constraint can change how the agent writes, qualifies, or justifies its conclusions more easily than it changes where the agent navigates. This creates a possible mismatch between evidential stance and evidence-gathering behavior.

Figure~\ref{fig:rq2-adj-boxplot} reports \emph{termination calibration}, \emph{navigation grounding}, and \emph{synthesis grounding} by model and condition. Here, the main condition-level shift is in termination calibration, which decreases under pressure (0.715 under natural versus 0.633 under budget pressure). Navigation grounding and synthesis grounding change more subtly across conditions. Under prior suppression, synthesis grounding decreases slightly, while navigation grounding remains similar or shifts marginally in the opposite direction.

At the action level, structural overview actions show the strongest prior-leaning navigation, with a grounding score of 0.464 compared to 0.827 for explore-code actions. Conclusion and memory actions show the strongest prior-leaning synthesis, with scores of 0.497 and 0.525. This asymmetry is consistent with prior suppression affecting synthesis-oriented output more strongly than navigation-oriented behavior. We revisit possible interpretations of this mismatch in Section~\ref{sec:dis-align-works}.

\begin{table}[t]
\centering
\caption{Automatic adjudication scores for \emph{navigation grounding} and \emph{synthesis grounding}. The table reports task-by-model means, with each cell showing navigation grounding/synthesis grounding. ``Overall'' is the task-level mean across models.}

\label{tab:rq2-auto-adj-task-prior-usage}
\footnotesize
\setlength{\tabcolsep}{3.5pt}
\renewcommand{\arraystretch}{1.10}

\begin{tabular}{
>{\raggedright\arraybackslash}m{0.18\linewidth}*{5}{>{\centering\arraybackslash}m{0.14\linewidth}}
}
\toprule
{Task} & {Overall} & {GPT-4.1} & {GPT-4o} & {DeepSeek-V3.2} & {Qwen3.5-397B} \\
\midrule
\emph{architecture analysis} & \mbox{0.691/0.558} & \mbox{{0.666}/0.481} & \mbox{0.751/0.804} & \mbox{0.674/0.400} & \mbox{0.681/{0.375}} \\
\addlinespace[0.35em]
\emph{ASR inference} & \mbox{\textbf{0.615}/\textbf{0.410}} & \mbox{\textbf{0.568}/\textbf{0.367}} & \mbox{\textbf{0.698}/\textbf{0.631}} & \mbox{\textbf{0.600}/\textbf{0.268}} & \mbox{\textbf{0.607}/0.370} \\
\addlinespace[0.35em]
\emph{feature implementation tracing} & \mbox{0.712/0.588} & \mbox{{0.695}/0.604} & \mbox{0.736/0.646} & \mbox{0.715/0.614} & \mbox{0.711/{0.389}} \\
\addlinespace[0.35em]
\emph{risk and code smell review} & \mbox{0.685/0.570} & \mbox{0.691/0.516} & \mbox{0.730/0.644} & \mbox{{0.658}/0.658} & \mbox{0.681/\textbf{0.333}} \\
\bottomrule
\end{tabular}
\end{table}

Table~\ref{tab:rq2-auto-adj-task-prior-usage} reports grounding scores by task family. \emph{ASR inference} is the most prior-shaped task on both dimensions (navigation: 0.615, synthesis: 0.410), consistent with a task that requires speculating beyond what static code reveals.

\Needspace{12\baselineskip}
\begin{finding}
\textbf{\textsc{signals}:} Lens~3 projects trajectory outcomes into adjudicated conclusion-level and epistemic-grounding scores (Figures~\ref{fig:rq2-scatter}--\ref{fig:rq2-adj-boxplot}; Table~\ref{tab:rq2-auto-adj-task-prior-usage}).
\textbf{\textsc{profile}:} Adjudicated quality is primarily model-characteristic and comparatively stable across conditions. The projection also reveals mismatches between dimensions, including cases where high grounding co-occurs with low task relevance (Figure~\ref{fig:rq2-scatter}). Under budget pressure, termination calibration decreases, while navigation and synthesis grounding shift more subtly (Figure~\ref{fig:rq2-adj-boxplot}).
\textbf{\textsc{boundary}:} These scores identify where trajectories align or misalign with the judged criteria, but they do not by themselves establish a causal mechanism for the observed mismatches; interpretation is deferred to Discussion (Section~\ref{sec:discussion}).
\end{finding}

\begin{figure}[p]
  \centering
  \includegraphics[width=0.95\linewidth]{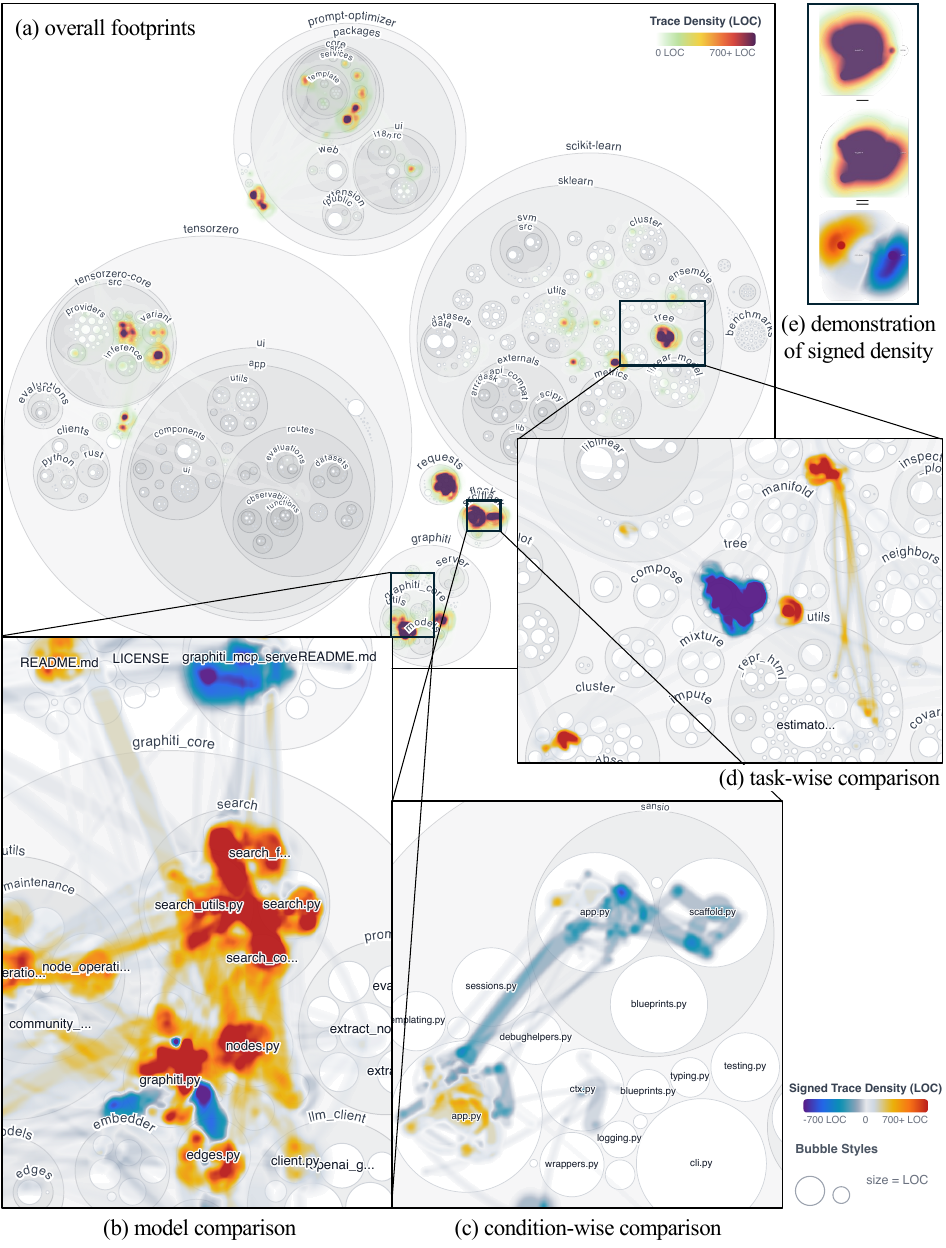}
  \caption{Repository footprints projected through trace density. Panel~(\subref{fig:rq2-footprint-a}) shows an overview of all six repositories with cumulative trace density across the full matrix. Panel~(\subref{fig:rq2-footprint-b}) compares models: signed density map of Qwen3.5-397B minus DeepSeek-V3.2 on Graphiti, showing divergent spatial attention on the same repository and task. Panel~(\subref{fig:rq2-footprint-c}) compares conditions: signed density map of Natural minus budget pressure on Flask, showing spatial contraction under constraint with locally intensified heat zones. Panel~(\subref{fig:rq2-footprint-d}) compares tasks: \emph{architecture analysis} minus \emph{feature implementation tracing} on scikit-learn, showing broad architectural exploration against narrow implementation focus. Panel~(\subref{fig:rq2-footprint-e}) provides the legend for signed density encoding.}
  \label{fig:rq2-footprint}
  \phantomsubcaption\label{fig:rq2-footprint-a}
  \phantomsubcaption\label{fig:rq2-footprint-b}
  \phantomsubcaption\label{fig:rq2-footprint-c}
  \phantomsubcaption\label{fig:rq2-footprint-d}
  \phantomsubcaption\label{fig:rq2-footprint-e}
\end{figure}

\subsubsection{Footprint: Spatial Divergence in the Repository World}\label{sec:res-rq2-lens-4}

Lens~4 (the codebase footprint lens) projects each trajectory into a spatial trace-density map over files and code regions, making repository contact directly comparable across runs. This evidence-layer projection reveals systematic differences in spatial attention across models, constraints, and tasks, while also preserving meaningful single-run variation.

Figure~\ref{fig:rq2-footprint} visualizes these footprint projections. Panel~(\subref{fig:rq2-footprint-a}) summarizes aggregate footprints across all six repositories and shows that large regions remain unvisited, with attention concentrated on a subset of structurally central files. Panel~(\subref{fig:rq2-footprint-b}) contrasts models on the same repository and task, showing that spatial attention can diverge markedly even when the experimental configuration is held fixed. Panel~(\subref{fig:rq2-footprint-c}) contrasts conditions and shows spatial contraction under budget pressure, with coverage narrowing to fewer files, while retained hotspots can intensify locally. Panel~(\subref{fig:rq2-footprint-d}) contrasts tasks and shows that architecture analysis induces broader coverage than feature implementation tracing, which concentrates on a smaller module neighborhood.

\begin{figure}[t]
  \centering
  \includegraphics[width=0.5\linewidth]{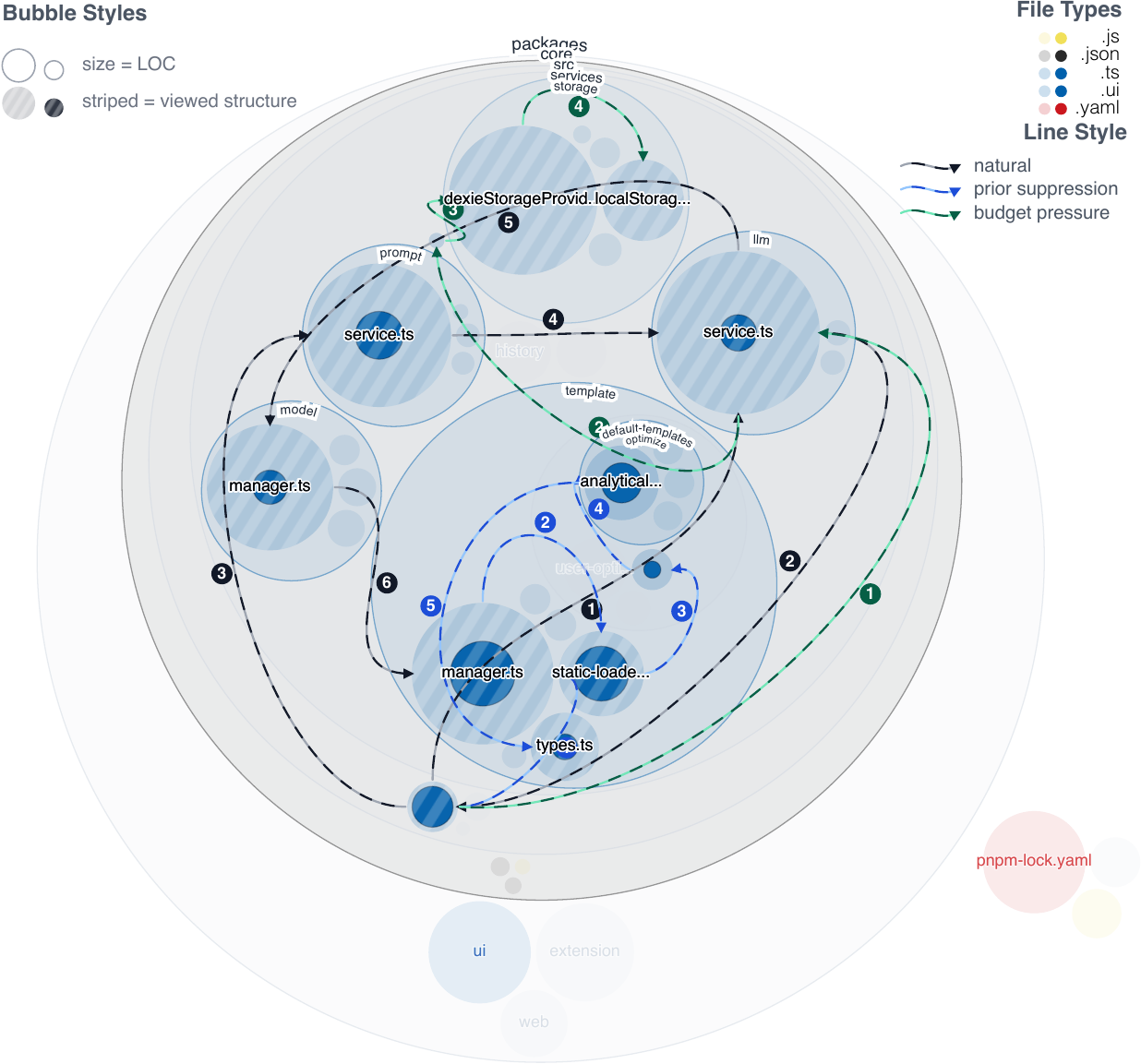}
  \caption{Under different conditions, even the similar prior suppression condition and natural condition are spatially different. They might have overlaps on files, but the order is unpredictable. Budget compression shows a more compressed amount of navigation in this view.}
  \label{fig:rq2-trajectory-divergence}
\end{figure}

In addition to these aggregate comparisons, the footprint projection remains informative at the run level. Figure~\ref{fig:rq2-trajectory-divergence} shows that, even for the same model, task, and repository, individual trajectories can follow different spatial routes through the codebase.

A boundary of the footprint projection is that it inherits artifacts from the underlying tool interface. For example, some configurations exhibit anomalous single-step intake (e.g., reads exceeding 1{,}000 LOC in one action), which can inflate local density independent of a sustained exploration strategy.

\begin{finding}
\textbf{\textsc{signals}:} Lens~4 projects repository contact as trace-density over files and code regions.
\textbf{\textsc{profile}:} The footprint projection distinguishes spatial attention patterns by model, shows coverage contraction under budget pressure with locally intensified hotspots, and reflects task-dependent coverage differences. It also preserves single-run variability, with different trajectories traversing different regions under the same configuration.
\textbf{\textsc{boundary}:} Density is shaped by the repository interface (e.g., occasional single-step large-LOC reads), so local hotspots should be interpreted as evidence of contact rather than as a direct measure of sustained effort.
\end{finding}

\FloatBarrier

\medskip

\noindent\textbf{Summary of RQ2.}\quad
The four observation lenses, read together, project a consistent pattern from different vantage points. Models carry distinct behavioral profiles that are visible in their journey shapes (lens~1), action grammars (lens~2), quality signatures (lens~3), and spatial footprints (lens~4). Budget pressure compresses and restructures these profiles in model-specific directions, but does not collapse conclusion quality. Prior suppression is invisible to the surface, temporal, and action-level lenses, and becomes detectable only through the adjudication lens, where it constrains synthesis more than navigation. Across all lenses, the same finding recurs at increasing resolution: trajectories diverge while conclusions converge. Different models, different conditions, and even different runs of the same configuration produce distinct journeys through the repository world, yet arrive at comparable understanding. The projection protocol makes this divergence visible and analyzable. What it cannot do is intervene. Whether an external intelligence can improve the trajectory rather than merely observe it is the question RQ3 addresses.

\subsection{RQ3: Does Intervention Improve or Expose a Boundary?}\label{sec:res-rq3}

\begin{figure}
    \centering
    \includegraphics[width=\linewidth]{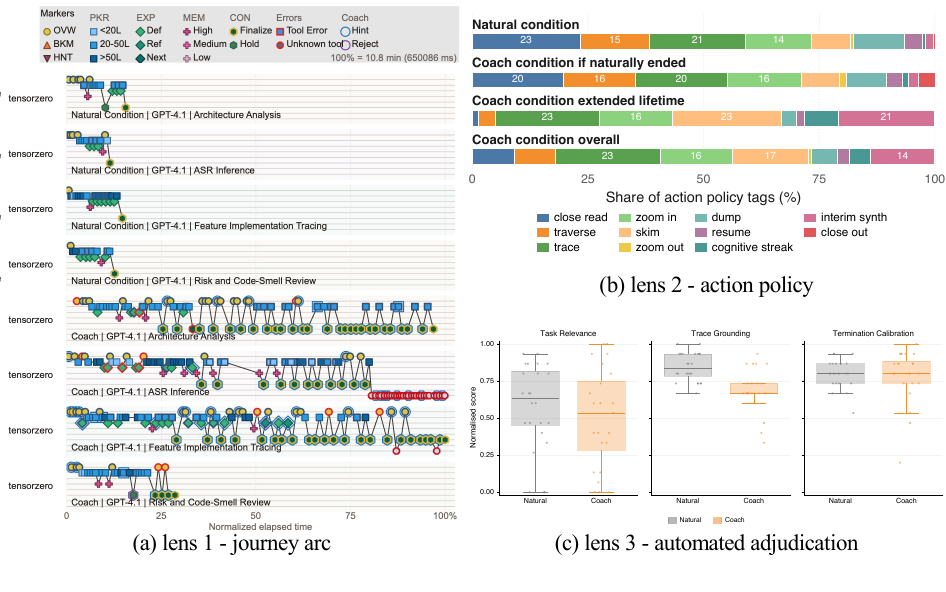}
    \caption{Coach-judge condition observed through three lenses, all on GPT-4.1. (a) Journey arcs comparing natural condition and Coach Condition on TensorZero across four task families. Natural trajectories terminate within the first quarter of the coach timeline. Coach trajectories extend substantially, with late-phase segments dominated by repeated attempts to conclude and coach rejections. (b) Action policy signatures split by lifetime regime. The natural-lifetime portion of coach trajectories closely resembles the natural condition. The extended-lifetime portion shifts toward close-out and cognitive-streak dominance, indicating a regime change from exploration to conclusion cycling. (c) Adjudication scores comparing Natural and Coach conditions. \emph{Conclusion grounding} drops under coaching; \emph{Task relevance} spreads wider with more extreme low scores; \emph{Termination calibration} remains stable.}
    \label{fig:rq3-lenses}
    \phantomsubcaption\label{fig:rq3-lenses-a}
    \phantomsubcaption\label{fig:rq3-lenses-b}
    \phantomsubcaption\label{fig:rq3-lenses-c}
\end{figure}

\begin{figure}
    \centering
    \includegraphics[width=0.7\linewidth]{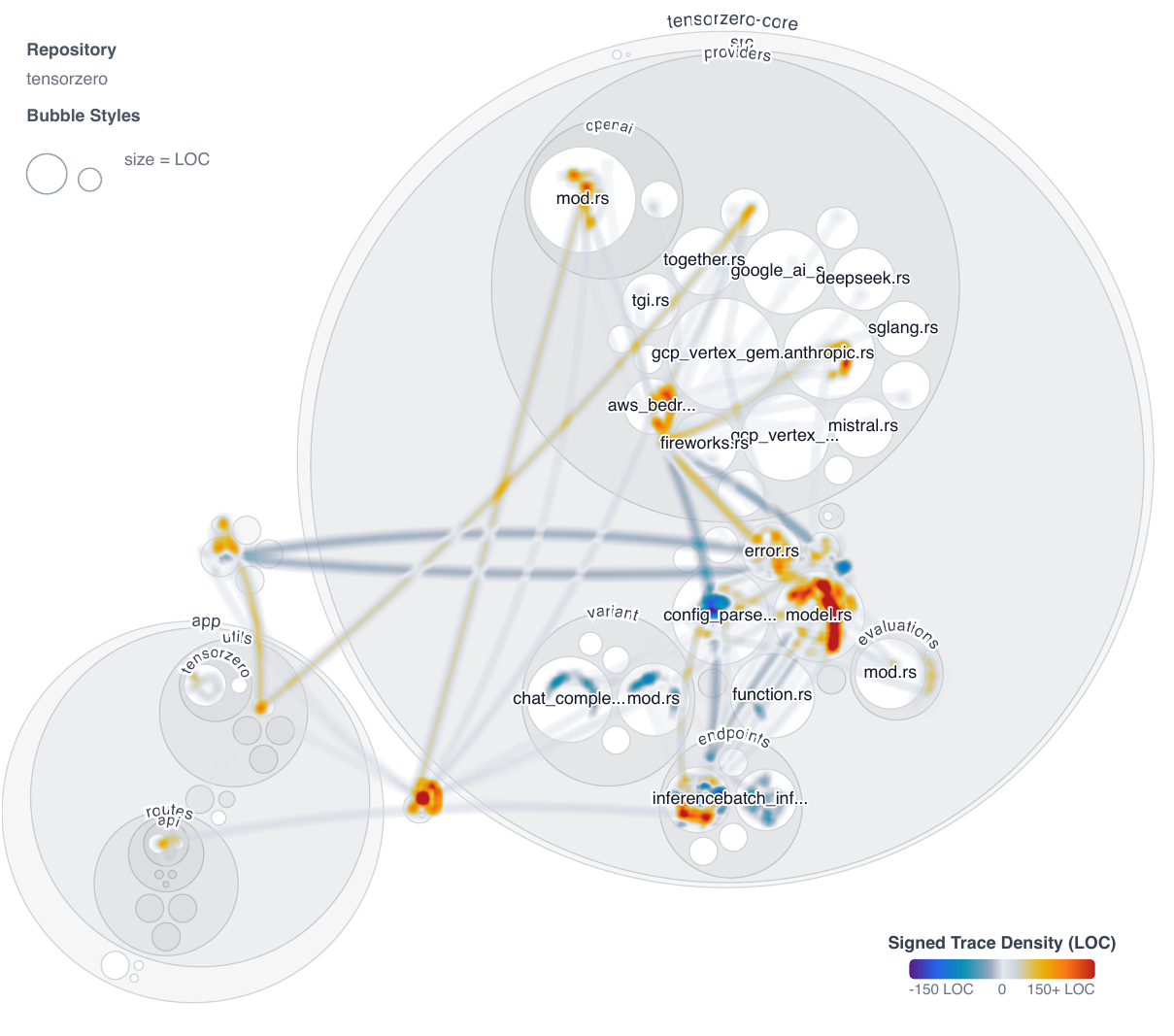}
    \caption{Signed trace density map comparing coach condition minus natural condition for GPT-4.1 on TensorZero. Orange and red zones indicate files where the coach condition reads more code; blue zones indicate files where the natural condition reads more. The map visualizes how spatial attention shifts under intervention, including dispersion across a broader set of files versus concentration on fewer files with deeper coverage.}
    \label{fig:rq3-footprint}
\end{figure}

RQ3 evaluates an intervention condition in which a judge panel and a coach are introduced into GPT-4.1's observation channel, yielding 24 additional trajectories across the same repositories and task families. We read this condition through the same lens stack as RQ2 (Figure~\ref{fig:rq3-lenses}, plus the footprint projection in Figure~\ref{fig:rq3-footprint}). At the level of these projections, coaching increases trajectory volume but is associated with a shift toward late-phase conclusion cycling and reduced conclusion grounding.

Relative to the natural condition, the coach condition generates substantially more surface activity: 2.5$\times$ more turns, 3.4$\times$ more files visited, and 3.2$\times$ more total time. Per-turn time remains nearly identical (5.50\,s vs.\ 5.32\,s), so the increase is primarily in trajectory length rather than per-turn duration. The amount of extension also varies by repository; some trajectories terminate earlier, while others (notably on TensorZero) extend to the 100-turn bound.

Inside Figure~\ref{fig:rq3-lenses}, panel~(\subref{fig:rq3-lenses-a}) shows that natural-condition trajectories on TensorZero terminate within the first quarter of the coach-condition timeline, whereas coach-condition trajectories extend across the full duration and exhibit late-phase segments dominated by repeated conclusion attempts and coach rejections. Panel~(\subref{fig:rq3-lenses-b}) shows the corresponding transition-profile split: the natural-lifetime portion of coach trajectories remains close to the natural condition, while the extended-lifetime portion shifts toward close-out (21\% of transitions) and cognitive streaks, with reduced close reading and traversal.

Panel~(\subref{fig:rq3-lenses-c}) reports adjudication outcomes. Under coaching, \emph{conclusion grounding} shifts from a tight natural-condition range around 0.85--0.95 to a lower median (around 0.68) with wider spread, and \emph{task relevance} becomes more dispersed with more extreme low scores. \emph{Termination calibration} remains similar between conditions, indicating that extended trajectories do not yield improved calibration signals under this projection.

Figure~\ref{fig:rq3-footprint} provides the footprint projection for the coach condition relative to the natural condition on TensorZero, making the intervention's spatial redistribution of repository attention visible.

\begin{finding}
\textbf{\textsc{signals}:} Under intervention, lens~1 exposes extended trajectory lifetime and late-phase segments dominated by repeated conclusion attempts and coach rejections (Figure~\ref{fig:rq3-lenses-a}); lens~2 splits transition profiles by regime and shows a shift in the extended lifetime toward close-out and cognitive-streak transitions (Figure~\ref{fig:rq3-lenses-b}); lens~3 reports adjudication changes in grounding and relevance (Figure~\ref{fig:rq3-lenses-c}); and lens~4 maps the corresponding spatial redistribution of repository attention (Figure~\ref{fig:rq3-footprint}).
\textbf{\textsc{profile}:} Coaching increases trajectory volume and extends runs, but the additional lifetime is associated with a transition-profile shift away from exploration and with lower, more variable conclusion grounding, while termination calibration remains similar between conditions.
\textbf{\textsc{boundary}:} In this apparatus, the intervention channel increases activity without yielding improved grounding signals under adjudication. This constitutes an empirical boundary for externally steered trajectories in our setting: additional interaction does not necessarily translate into stronger repository-grounded conclusions.
\end{finding}

\section{Discussion}\label{sec:discussion}

\subsection{Situating Our Findings in the Broader Literature}\label{sec:dis-align-works}

Ada's apparatus was designed for behavioral characterization rather than large-scale causal inference. Its findings are structurally reliable, grounded in 384 trajectories across four models, six repositories, and four constraint conditions, but they are best interpreted alongside related work. Prior studies address adjacent phenomena, including agent efficiency, diversity of reasoning paths, chain-of-thought faithfulness, epistemic constraints, and multi-agent coordination failures. Table~\ref{tab:findings-in-context} situates Ada's findings in this landscape by pairing each observation with the closest evidence in the literature and its citation. In several cases, Ada recovers a known effect at trajectory resolution; in others, it provides process-level characterization that outcome-centric studies do not expose.

{\footnotesize
\setlength{\tabcolsep}{4pt}
\renewcommand{\arraystretch}{1.15}
\begin{longtable}{@{}
    p{0.32\textwidth}
    p{0.52\textwidth}
    p{0.08\textwidth}
@{}}
\caption{Ada's observed patterns aligned with prior work. Each row pairs one observation from Ada's projection study with the closest evidence in the community literature and its citation.} \label{tab:findings-in-context} \\
\toprule
Observed in Ada & Closest findings in prior works & Citation \\
\midrule
\endfirsthead
\toprule
Observed in Ada & Closest findings in prior works & Citation \\
\midrule
\endhead
\bottomrule
\endfoot
\bottomrule
\endlastfoot

Different models traverse different repository regions yet arrive at comparable conclusions & Complex reasoning tasks admit multiple diverse reasoning paths that reach the same correct answer & \cite{wang2022selfconsistency} \\
& Different SWE agents resolve very different sets of issues despite having similar resolve rates; the large action space of agents inevitably produces solution diversity & \cite{zhang2024dei} \\
\midrule

Conclusion quality is model-characteristic and largely condition-stable; the outward surface of behavior is a weak predictor of synthesis quality & Agentic system evaluation has been mostly outcome-centric, overlooking intermediate steps and masking recurrent inefficiencies; process-centric metrics are needed to assess quality independent of final success & \cite{liu2025graphectory} \\
& LLM-generated reasoning chains are often not causally linked to the final answer; perturbing intermediate steps may not change predictions, making explanations essentially post-hoc & \cite{arcuschin2025cotfaithfulness} \\

\addlinespace[0.35em]
\midrule

Budget pressure compresses trajectory volume while conclusion quality remains largely stable, suggesting unrealized efficiency in unconstrained trajectories & Longer reasoning / extra turns show diminishing returns: stopping earlier can maintain comparable accuracy, and max-turn trajectories are unlikely to yield substantial gains. & \cite{zhou2026overthinking,yang2024sweagent} \\
& Inefficiency and budget interaction: redundant steps waste compute without improving outcomes, and budget effectiveness depends on scaffold--LLM synergy (with failed attempts consuming disproportionate tokens). & \cite{gandhi2025sweprm,fan2025sweeffi} \\
\pagebreak

Wall-clock time gaps in API-mediated trajectories create an attribution boundary for temporal projections; pauses cannot be interpreted uniquely as additional agent work & Latency variability in LLM serving can confound time-based measures, and gray-failure dynamics can manifest as partial slowdowns without fail-stop errors & \cite{li2025continuum,huang2017gray} \\
\addlinespace[0.35em]
\midrule

Prior suppression constrains synthesis more than navigation; agents adjust written claims more than evidence-gathering behavior, consistent with epistemic specification gaming & Specification gaming occurs when agents satisfy the literal objective but not its intent; LLMs generalize from simple forms of gaming to more sophisticated compliance failures that are nontrivial to remove & \cite{denison2024sycophancy} \\
& LMs often ignore provided context when it conflicts with pre-existing parametric memory; navigational behavior draws on ingrained priors that prompt-level instructions cannot easily override & \cite{augenstein2026parametric} \\
\addlinespace[0.35em]
\midrule

External coach override reduces trace grounding while leaving natural termination calibration intact; added interaction increases activity but degrades outcomes & LLM teams consistently underperform their best individual member by 8--38\%; the failure is not identification but leveraging, because teams average expert and non-expert views rather than appropriately weighting expertise & \cite{pappu2026multiagentexperts} \\
& Multi-agent integration success degrades monotonically as shared specification decreases; a substantial coordination tax persists even with full specifications & \cite{chaconsartori2026specgap} \\
\addlinespace[0.35em]
\midrule

Coach fuses multi-judge critiques into a single instruction; this lossy aggregation can discard the strongest signal and obscure responsibility boundaries & Diversity only helps with good aggregation: synthesis can lose to single-model baselines, while selection/competition can beat synthesis & \cite{maryanskyy2026selectionbottleneck} \\
& Role-specialized pipelines can silently propagate errors without auditable handoffs; critique is not independently verified before reuse & \cite{barrak2025traceability} \\
\end{longtable}
}

Across the seven rows, we see a consistent pattern. Our experiments are small in scale, but the composed methods and lenses let us project trajectories from multiple views. The mechanisms we observe align with prior work at a larger scale, including inefficient exploration, divergent reasoning paths with similar endpoints, output shifts that exceed behavior change, and coordination loss in multi-agent systems. Ada adds structure by making these effects visible in the layers of narrative, mechanism, and evidence. It also pinpoints specification gaming at the navigation-to-synthesis boundary and shows how oversight can push agents from grounded exploration into conclusion cycling. The projection protocol is what makes these dynamics visible.

\subsection{Making the Code-Understanding Journey Observable}

The contribution of this work is observational. By treating trajectories as empirical objects, the apparatus and lenses make repository-level code understanding comparable as a process: what evidence is inspected, how navigation and revisitation unfold, when intermediate understanding is externalized, and how stopping is calibrated. These comparisons identify projected behavioral differences under controlled launch conditions. They do not by themselves establish causal mechanisms inside the model.

This framing also bounds the interpretation of the launching conditions. Differences across conditions should be read as behavior under controlled changes in guidance within a fixed instrument, not as intrinsic properties of the underlying models or as guarantees of transfer to other software environments.

More broadly, Ada provides a reusable observation design for SWE agents. It projects externalized behavior through explicit lenses to support responsible, process-level claims without assuming direct access to internal cognition.

\subsection{What Can Travel Beyond Ada}

The preceding subsections situate Ada's findings in the literature and clarify the observational contribution of the apparatus and lenses. What can travel beyond Ada is therefore not the specific behavioral profiles reported here, but the observation logic that makes such profiles readable.

The method is transferable only with design work. Other researchers should not treat Ada as a universal template that can simply be copied across agents, tools, or domains. What can travel is the logic of the apparatus. A future study can define a bounded software world, specify meaningful task families, expose tools that shape the agent's contact with the environment, and design lenses that match the phenomenon under observation. Under that condition, trajectory-centered projection can become a practical method for studying SWE agents across different contexts. The findings here are not the end of that program. They are an argument that such a program is possible.

\section{Threats and Limitations}\label{sec:threats-limits}

The limitations of this study are part of the methodological conditions under which trajectory-level observation becomes possible. Ada was designed to project repository-level code understanding from behavioral traces, while treating the underlying models as black boxes. This gives access to how an agent moves, reads, records, revises, and terminates inside a bounded repository world. It does not give access to why the model internally preferred one hypothesis, file, or conclusion over another. The three limitations below clarify the epistemic boundaries within which the reported findings hold.

\subsection{Apparatus Dependence}

Any attempt to study an SWE agent in action requires deciding what world the agent enters, what actions it can take, what feedback it receives, and what counts as completion. These decisions constitute the observation environment, and they shape the trajectories that result. Ada is therefore not a neutral window onto SWE agent behavior. Its task openings, prompt scaffold, tool surface, repository interface, and termination mechanism all define what becomes observable, and what remains outside view. The four experimental conditions are prompt-level and interaction-policy perturbations over a single scaffold. They do not change the underlying toolset or runtime architecture, which strengthens cross-condition comparison but means that observed behavioral differences depend on how the perturbations were designed. Model selection introduces a parallel dependence. The four foundation models used for trajectory generation, the post-hoc evaluation models, and the runtime coach-judge layer belong to different parts of the experimental stack, and the study cannot claim that the same profiles would appear under all model families or all prompt formulations.

Our design addresses this dependence by keeping the core scaffold stable across conditions and changing only specific forms of guidance or decision pressure. The resulting trajectories should be read as behavior under Ada's apparatus. They speak to this setting, and the apparatus is designed for structured observation across conditions, not for claims about SWE agents in the abstract. Precisely because this dependence is unavoidable, making it explicit is a methodological contribution. The community benefits from observation environments whose assumptions are inspectable, rather than from claims that appear apparatus-free.

\subsection{Scope of Repository-Level Code Understanding}

Code understanding is one phase of software engineering work, and this study deliberately isolates it. The task families ask the agent to understand, trace, infer, and explain code. They do not ask it to patch bugs, run tests, manage issue histories, negotiate requirements, or maintain software over time. This scope removes patch success as the organizing endpoint and places code understanding itself at the center of observation. It also means that the findings should not be transferred directly to repair-centered SWE agent settings.

The repository sets bounds on the study in a second way. Six repositories span compact libraries, classic frameworks, scientific software, and recent AI-infrastructure systems, providing a structurally heterogeneous terrain for observation. Heterogeneity, however, is not coverage. Six repositories cannot represent the full space of software systems, organizational practices, programming languages, or development contexts. The scale difference across the set, from compact repositories to systems exceeding 100K LOC, helps stress the apparatus but also means that repository complexity may interact with model behavior and task type in ways this study can only begin to characterize. These boundaries are intentional. They define the experimental field within which the observation method can be demonstrated and assessed, while leaving the question of broader applicability to future work with expanded repository sets and task designs.

\subsection{Lens Dependence}

Behavioral profiles are projections, not the agent's raw properties. Each lens makes certain aspects of the trajectory visible while leaving others outside the view. Surface statistics, tool-use sequences, grounding judgments, prior-usage assessments, and termination evaluations each project a partial reading. Different lenses could emphasize different phenomena, such as uncertainty management, hypothesis revision, semantic compression, or interaction with repository topology. This partiality is inherent to the method, and it carries specific epistemic consequences that the reported findings cannot fully resolve.

One consequence concerns convergence. When trajectories diverge across models, but conclusions converge, the observation lenses cannot definitively distinguish genuine understanding from training-data recall or from task conditions that admit only a narrow range of acceptable answers. Prior suppression changes the evidential stance of the launch, but it operates at the prompt level and cannot guarantee the absence of parametric familiarity. Honest uncertainty about what convergence means is part of the current result.

A second consequence concerns the automatic adjudication lens specifically. This lens uses LLM-based judges to evaluate LLM-produced trajectories. The design mitigates circularity through model separation, anonymization, and multi-model voting, but the structural condition remains. Judges and agents share overlapping training-data distributions, so agreement among judges may reflect shared priors rather than independent validation. The alignment between sampled human evaluation and judge scores provides partial reassurance, but it does not eliminate the concern.

These limitations point toward the same methodological reminder. A trajectory does not interpret itself. Responsible study of SWE agent behavior requires explicit lenses whose assumptions can be inspected, reused, criticized, and redesigned. The set $\mathcal{L}$ demonstrated in this paper is open and minimal. Its value lies in showing that structured observation yields distinguishable behavioral signals, and in making the instruments available for the community to extend and refine.

\section{Related Works}\label{sec:related}

\subsection{Trajectory-Aware Evaluation of Software and Tool-Using Agents}

Existing work on software agents has increasingly moved beyond final-answer evaluation toward process-aware analysis. Early evaluations of coding agents often centered on whether a task was completed, a patch passed tests, or a benchmark item was solved. More recent work treats the intermediate trajectory as an empirical object in its own right. Studies of thought-action-result logs, diagnostic toolkits for agent trajectories, and trajectory-level evaluation frameworks examine how agents search, read, call tools, recover from errors, and terminate across multi-step processes \cite{bouzenia2025trajectories,ou2025agentdiagnose,kim2025beyondfinal}. This shift is important because long-horizon agent behavior cannot be adequately understood from the final output alone. A correct result may be reached through inefficient, fragile, or poorly grounded routes, while an incorrect result may still contain informative evidence about where the agent's process broke down. In this sense, trajectory-level analysis has become a necessary complement to outcome-level evaluation.

\subsection{Tool-Mediated Interaction and Repository Environments}

Within software engineering, this process-aware turn has been especially visible in work on repository-level agents. Recent benchmarks and datasets place agents inside realistic code environments, where success depends on locating relevant files, interpreting dependencies, maintaining state across turns, and handling setup or validation constraints \cite{gautam2025refactorbench,yang2025swesmith,ni2025gittaskbench,rashid2025swepolybench}. These studies make clear that software agents do not operate over isolated snippets. They work inside repository worlds composed of files, symbols, dependency relations, build structures, tests, and developer tools. The resulting behavior is therefore not only a property of the model. It is co-produced by the model, the task opening, the action surface, and the feedback returned by the environment.

A parallel line of agent research has studied tool-mediated interaction more generally. Tool-use benchmarks and surveys emphasize that agents must select actions, interpret stateful feedback, manage tool dependencies, and adapt over multiple turns rather than merely produce a single response \cite{lu2025toolsandbox,yao2024taubench,wang2025stateful}. This literature supports a central assumption of our apparatus. The tool interface is not a neutral channel through which an agent simply accesses information. It shapes what can be observed, what can be attempted, and what can become part of the trajectory. For SWE agents, this point is especially consequential because code-reading tools, structural overviews, reference tracing, memory mechanisms, and conclusion actions define the agent's mode of contact with the repository. A trajectory should therefore be read as a tool-mediated encounter between an agent and a software world.

\subsection{Repository-Level Code Understanding and Software Comprehension}

Repository-level code understanding provides the second foundation for this work. Modern LLM-based code research has moved from single-function or single-file tasks toward cross-file completion, repository-level editing, dependency understanding, context retrieval, and code question answering \cite{liu2023repobench,ding2023crosscodeeval,bairi2024codeplan,du2025dependeval,peng2025sweqa}. These works show that repository understanding is not merely a matter of giving the model a larger context window. It requires selecting relevant evidence, following dependencies across files, distinguishing central from peripheral code, and forming a coherent account of how parts of a system interact. This view is close to the problem setting of our study, but our focus differs. We do not ask whether an agent can complete, edit, or answer a repository task as an endpoint. We ask how its code-understanding journey can be made observable while it moves through the repository.

The task families used in our apparatus also connect to long-standing software engineering research. Architecture analysis and architecture reconstruction have studied how high-level design structure can be inferred from implementation artifacts \cite{murphy1995reflexion,ducasse2009architecture}. Work on architecture evaluation and architecturally significant requirements explains why some design decisions and implicit requirements have disproportionate structural impact \cite{kazman2000atam,chen2013asr}. Feature location research studies how user-visible behavior can be traced to source-code units, while inspection and code-review research grounds risk-oriented reading in established maintenance practices \cite{eisenbarth2003locating,fagan1976inspections,bacchelli2013codereview}. These traditions matter because they prevent repository-level code understanding from being treated as a new capability invented by LLMs. Architecture reasoning, requirement inference, feature tracing, and risk review are established forms of software comprehension. Our work brings these forms into an apparatus for agent trajectories.

\subsection{From Process Evaluation to Projected Observation}

Taken together, these bodies of work establish three premises. First, final outcomes alone are too narrow for understanding long-horizon agent behavior. Second, agent trajectories are shaped by the tools and environments through which agents act. Third, repository-level code understanding is a cross-file, evidence-selective, and historically grounded software comprehension problem. However, existing work still leaves a methodological gap. Trajectory-aware studies often use process traces to diagnose failure, improve repair performance, train agents, or benchmark execution. Repository-level code benchmarks often retain task success as the organizing endpoint. Classical software comprehension explains what kinds of understanding matter, but it does not address how an LLM-based SWE agent's understanding process can be observed from outside.

This paper occupies that gap. We construct Ada as a scoped SWE agent apparatus for observing repository-level code-understanding trajectories in a wild-but-bounded software world. Ada is not intended as a benchmark agent, and its trajectories are not treated as transparent access to internal cognition. Instead, each trajectory is recorded as a recoverable observation record and projected through analysis lenses that make behavioral and epistemic profiles discussable. This positioning distinguishes our work from outcome-centered SWE agent evaluation and from repair-oriented trajectory diagnosis. Prior work shows that agent trajectories useful for evaluation, diagnosis, correction, and scaling. We use trajectories to make repository-level code-understanding journeys observable.

\section{Conclusions}\label{sec:conclusion}

This paper introduced \emph{Ada}, a scoped SWE agent apparatus for observing repository-level code-understanding as a trajectory-level empirical object. By combining a bounded-yet-wild repository world, a read-only tool-mediated interface, and a reusable suite of observation lenses, the method makes agent exploration, synthesis, and stopping behavior comparable across models and controlled perturbations without treating internal cognition as directly accessible.

Across 408 recorded trajectories, our projection study surfaced several stable signals. Budget pressure consistently compressed activity volume while leaving conclusion quality largely intact, revealing unrealized efficiency in unconstrained trajectories. Prior suppression altered the \emph{stance} of synthesis more than the \emph{shape} of navigation, suggesting that prompt-level evidential constraints are easier for agents to satisfy in their written claims than in their evidence-gathering behavior. Finally, injecting an external coach-judge sidecar increased interaction and activity but could degrade grounding, highlighting a coordination tax at the boundary where a solo trajectory becomes a multi-intelligence process.

Taken together, these results argue for a methodological shift: progress on SWE agents should be evaluated not only by endpoints, but by projected observation of the process that produces them. Ada is not a universal template; it is an observation logic that can be redesigned for other domains, tool surfaces, and task families. Future work should expand the repository set and task openings, add lenses for uncertainty management and hypothesis revision, strengthen validation via human and non-overlapping automated judges, and connect trajectory-level signals to downstream engineering outcomes (e.g., repair success) while preserving the interpretability that trajectory-centered observation provides.

\backmatter

\section*{Declarations of Conflict of Interest}

The authors declared that they have no conflicts of interest in this work.

\noindent

\section*{Data Availability}

The apparatus source code, all 408 recorded trajectories, observation lens outputs, analysis scripts, and evaluation prompt templates are publicly available as a replication package at \url{https://doi.org/10.5281/zenodo.20337565}. The package includes raw trajectory data, pre-computed projection results for all five lenses, and the scripts that produce every figure and table reported in this paper.

\section*{Usage of AI Tools}

AI-assisted tools (Claude - Anthropic, Grammarly, and Overleaf's AI assistant features) were used during manuscript preparation for proofreading and readability improvements. Analysis scripts for trajectory statistics and visualization code for scientific figures were written with AI assistance under author-specified instructions and reviewed by the authors. The AI tools were not used to generate research ideas, design the study methodology, or produce images directly used in the paper. The authors take full responsibility for the content of this work.

\bigskip
\begin{flushleft}%
Editorial Policies for:

\bigskip\noindent
Springer journals and proceedings: \url{https://www.springer.com/gp/editorial-policies}

\bigskip\noindent
Nature Portfolio journals: \url{https://www.nature.com/nature-research/editorial-policies}

\bigskip\noindent
\textit{Scientific Reports}: \url{https://www.nature.com/srep/journal-policies/editorial-policies}

\bigskip\noindent
BMC journals: \url{https://www.biomedcentral.com/getpublished/editorial-policies}
\end{flushleft}

\begin{appendices}

\section{Appendix: Experiment Pipeline}\label{app:experiment-pipeline}

This appendix documents the experimental apparatus behind Ada and the Coach-Judge condition. It is intended as replication-oriented scientific reporting: it describes what the apparatus asks the agent to do, what information the agent can observe, how the experimental conditions modify the agent's operating stance, and how the coach-judge process enters a trajectory. It does not report behavioral findings.

\subsection{Ada Apparatus}

\subsubsection{Prompt Scaffold}

Ada is launched as a read-only code-understanding agent. The prompt scaffold gives it an identity, a constrained action protocol, a repository context, and a persistent state for accumulating understanding across turns. The scaffold is designed to make each trajectory observable: every turn contains stated reasoning, one selected action, and the environment's response.

\begin{figure}
  \includegraphics[width=\linewidth]{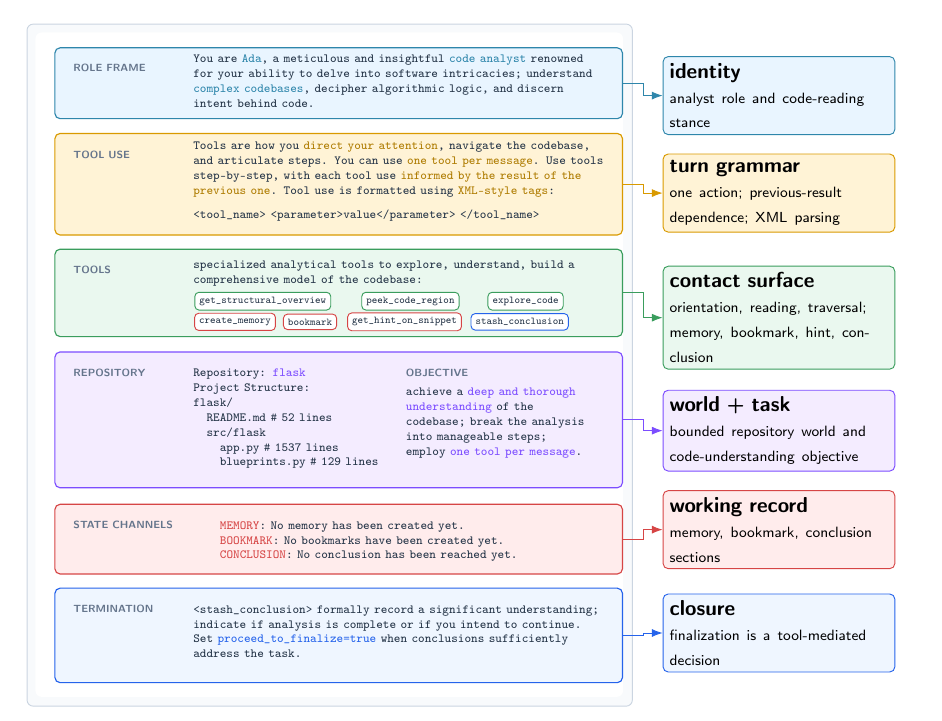}
  \label{fig:ada-prompt-scaffold}
  \caption{Prompt scaffold of Ada as a repository-level code-understanding agent, organized and abstracted by functional components. The figure compresses the role frame, one-tool-per-turn protocol, tool interface, repository context, state channels, and explicit termination rule into a prompt-listing view.}
\end{figure}

Two exact prompt excerpts matter for interpreting the trajectory design. The first is the one-action-per-turn constraint.

\begin{verbatim}
You must only use one tool per interaction, and your analysis 
must build up step-by-step from earlier results.
\end{verbatim}

The second is the termination contract. A trajectory ends only when the agent uses the conclusion tool and explicitly marks the analysis as ready to finalize.

\begin{verbatim}
- proceed_to_finalize: (required) A boolean value (true or false).
  - Set to true if you believe this conclusion ... 
    sufficiently addresses the code reading task, 
    and you wish to finalize your analysis.
\end{verbatim}

The prompt, therefore, turns code understanding into a bounded sequence of observable choices. The agent is not asked to answer immediately from prior knowledge. It is asked to reason, inspect the repository through the available tools, record intermediate understanding, and terminate only when it believes the task has been sufficiently addressed.

\subsubsection{Task Setup Prompts}

The system-side scaffold was paired with a short task block that specified what kind of repository understanding Ada should produce. These task prompts varied by Task Family, while the experimental conditions only changed Ada's operating stance around the same underlying task. The prompts below reproduce the user-facing task setups in paper-facing terminology rather than internal run identifiers.

\paragraph{Architecture Analysis.}
{\scriptsize
\begin{quote}
\ttfamily
GLOBAL GOAL: Analyze the provided codebase to determine its overall architecture and architectural style. You should identify the major components, understand how they are organized, and describe the primary patterns of interaction between them. Your analysis should conclude with a summary of the architectural design.

Constantly viewing the structure of code might be helpful. And creating memory on important logical and data flow will help you in concluding the architecture design.

You can start now.
\end{quote}
}

\paragraph{ASR Inference.}
{\scriptsize
\begin{quote}
\ttfamily
GLOBAL GOAL: Analyze the provided codebase to identify its Architecturally Significant Requirements (ASRs). Investigate the code for evidence of design decisions that satisfy implicit requirements such as performance, security, scalability, or reliability. Your analysis should conclude with a list of inferred ASRs, each supported by specific code evidence.

Constantly exploring the reference of code pieces might be helpful. Feel free to get a hint from the system.

You can start now.
\end{quote}
}

\paragraph{Feature Implementation Tracing.}
{\scriptsize
\begin{quote}
\ttfamily
GLOBAL GOAL: Your objective is to locate and describe the complete implementation of the target feature. Trace this feature from its most likely entry point through its core logic, data handling, and any interactions with other modules, concluding with a summary of the end-to-end implementation path.

If you get stuck, try bookmarking or getting a hint from the system. Constantly getting structure overviews of files is recommended. Verify your memories and conclusions before stopping.

You can start now.
\end{quote}
}

For this Task Family, the repository-specific target feature was substituted into the prompt. The six targets were request routing and dispatching in Flask, session management and cookie persistence in Requests, the \texttt{DecisionTreeClassifier} implementation path in Scikit-learn, conditional edge traversal in Graphiti, the core prompt-optimization workflow in Prompt Optimizer, and the streaming inference pipeline in TensorZero.

\paragraph{Risk and Code-Smell Review.}
{\scriptsize
\begin{quote}
\ttfamily
GLOBAL GOAL: Act as a code reviewer and systematically analyze the provided codebase to identify potential risks, comprehension difficulties, or code smells. Your analysis should result in a small list of these issues, with each issue referencing the specific location in the code where it was found.

Use bookmarks to flag code snippets that appear to be risky. You might want to scan the code files, so using peek region or exploring the code might be helpful.

You can start now.
\end{quote}
}

These task blocks were short by design. They specified the repository-understanding objective and a small amount of task-local guidance, while the scaffold and condition overlays controlled the action protocol, memory surfaces, termination rule, and evidential stance.

\subsubsection{Tool Surface}

The seven tools define Ada's contact surface with the repository and with its own accumulating state. Table~\ref{tab:appendix-tool-signatures} describes them by functional role rather than internal format. The navigation tools expose repository evidence; the cognitive tools externalize intermediate reasoning, unresolved questions, clarification requests, and final synthesis.

{\footnotesize
\begin{longtable}{@{}
  >{\raggedright\arraybackslash}p{0.15\textwidth}
  >{\raggedright\arraybackslash}p{0.12\textwidth}
  >{\raggedright\arraybackslash}p{0.34\textwidth}
  >{\raggedright\arraybackslash}p{0.30\textwidth}
@{}}
\caption{Ada tool signatures as exposed in the prompt schema.}\label{tab:appendix-tool-signatures}\\
\toprule
Tool & Group & Parameters & Function in trajectory \\
\midrule
\endfirsthead
\toprule
Tool & Group & Parameters & Function in trajectory \\
\midrule
\endhead
\bottomrule
\endfoot
\texttt{peek\_code\_region} & Navigation & \texttt{file\_path: string}; \texttt{start\_line\_number: number}; optional \texttt{end\_line\_number: number}. & Reads exact line ranges and records line-level file access. \\
\texttt{get\_structural\_\newline overview} & Navigation & \texttt{file\_path: string}; optional \texttt{logical\_block\_name: string}. & Calls the structure summarizer for a file or named block and records structural access. \\
\texttt{explore\_code} & Navigation & \texttt{action\_type: go\_to\_definition | find\_all\_references | go\_to\_next\_block}; \texttt{current\_file\_path: string}; \texttt{current\_line\_number: number}; conditional \texttt{symbol\_name: string}. & Uses VS Code definition/reference providers or retrieves the next local block. \\
\texttt{create\_memory} & Cognitive & \texttt{memory\_title: string}; \texttt{memory\_content: string}; optional \texttt{confidence\_level: high | medium | low}. & Stores intermediate hypotheses or synthesized understanding for later prompt reinjection. \\
\texttt{bookmark} & Cognitive & \texttt{file\_path: string}; \texttt{line\_number: number}; \texttt{note: string}. & Marks a code location for later attention. The Cline-oriented prompt names the note field \texttt{question\_or\_note}; the Copilot schema names it \texttt{note}. \\
\texttt{get\_hint\_on\_ \newline snippet} & Cognitive & \texttt{file\_path: string}; \texttt{code\_range: string}; \texttt{description\_of\_confusion: string}; optional \texttt{what\_i\_tried\_or\_think: string}. & Requests a bounded local hint for a confusing snippet. The hint generator is instructed not to answer the full task. \\
\texttt{stash\_conclusion} & Cognitive / termination & Optional \texttt{title: string}; \texttt{content: string}; \texttt{scope\_description: string}; \texttt{proceed\_to\_finalize: boolean}. & Records a major or final conclusion. Finalization occurs when \texttt{proceed\_to\_finalize} is true, corresponding to $u^{*}$ in Equation~\ref{eq:swe-agent}. \\
\end{longtable}
}

This tool surface is intentionally narrower than a full developer environment. Ada can inspect, navigate, remember, bookmark, ask for local hints, and conclude. It cannot edit files or execute arbitrary exploratory routines. This keeps trajectories centered on repository understanding rather than repair or code generation.

\subsubsection{Condition Variants}

The four main experimental conditions preserve the same task set, repositories, tool surface, and read-only boundary. They differ only in the stance imposed on Ada while it conducts the trajectory.

{\footnotesize
\begin{longtable}{@{}>{\raggedright\arraybackslash}p{0.10\textwidth}>{\raggedright\arraybackslash}p{0.53\textwidth}>{\raggedright\arraybackslash}p{0.26\textwidth}@{}}
\caption{Condition-level modifications to Ada's operating stance.}\label{tab:appendix-condition-variants}\\
\toprule
Condition & What changes & Rationale in the experiment \\
\midrule
\endfirsthead
\toprule
Condition & What changes & Rationale in the experiment \\
\midrule
\endhead
\bottomrule
\endfoot
Natural Condition & Uses the base scaffold (role, tool definitions/examples, tool-use guidelines, repository context, accumulated memory/bookmarks/conclusions, objective, notes/history, and the current task). No additional pressure or evidential constraint is added. & Establishes the unperturbed reference trajectory for the same agent, tasks, repositories, and tools. \\
Budget Pressure & Adds an explicit exploration budget: each turn is framed as consuming budget, so Ada must decide what not to inspect. A budget-status block is shown before the objective (default target: 12 exploration turns, unless overridden). Budget reminders are injected as the remaining budget drops. & Tests how the trajectory reorganizes when exploration space is scarce while the tool surface remains unchanged. \\
Prior Suppression & Adds an evidence-first stance: prior/parametric familiarity is treated as weak intuition rather than evidence. When drawing conclusions, Ada must ground them in repository-specific observations (e.g., files, symbols, code paths, or tool-observed details) and clearly separate hypotheses from observed evidence. & Tests whether the agent can separate likely prior familiarity from evidence gathered inside the current repository. \\
Combined Condition & Applies both constraints: budget-aware exploration plus evidence-first grounding. Ada is asked to maximize task-relevant certainty under the remaining budget while avoiding unsupported prior-driven conclusions. & Tests the joint constraint of scarce exploration and stricter evidential grounding. \\
\end{longtable}
}

Budget Pressure has both an initial prompt effect and a turn-time feedback effect. The initial prompt defines the budget as a real task constraint. During the trajectory, budget reminders are inserted at coarse milestones: around half the budget remaining, quarter budget remaining, three turns remaining, and the final turn. These reminders shift from narrowing the search space to consolidating and concluding from the strongest available evidence.

Prior Suppression does not attempt to remove prior knowledge from the model. Instead, it changes the evidential rule of the task: prior familiarity may suggest hypotheses, but the reported conclusion should be justified by what Ada has observed in the repository during the trajectory.

\subsubsection{Model Runtime and Context Handling}

Several runtime settings matter for interpreting Ada's trajectories. For the user-configurable trajectory-generation models, Qwen3.5-397B-A17B and DeepSeek-V3.2, the decoding temperature was fixed at 0.2. For the GitHub Copilot default models, GPT-4o and GPT-4.1, this parameter was not exposed by the runtime, so those trajectories inherit the provider's default behavior. Model reasoning mode was turned off in the Ada apparatus because the ReAct-style agent design already includes a visible thinking segment as part of the trajectory itself; enabling an additional hidden reasoning mode would have mixed two different reasoning surfaces inside the same observation record.

The context window was set to the largest capacity available for each model. When prompt assembly reaches that capacity, truncation is handled by the editor runtime's internal priority logic rather than by a study-specific manual rule. The system prompts were given higher priority, while accumulated navigation-turn messages were lower priority. As a result, when context overfill occurs, older or lower-priority trajectory material can be dropped, and some trajectories may therefore show goal drift that is partly induced by context saturation.

\subsection{Coach-Judge Pipeline}\label{app:coach-judge-pipeline}

\subsubsection{Pipeline Architecture}

The Coach-Judge condition adds a second reasoning process to the trajectory. Ada still performs one tool-mediated action per turn. After a turn is completed, the coach-judge process receives the completed turn and updates its trajectory record. Specialized judges then evaluate the turn from different perspectives, and a coach decides whether the next observation should remain unchanged or include intervention.

The functional flow is:

\begin{quote}
\small
Completed Turn \(\rightarrow\) Trajectory Summary \(\rightarrow\) Judge Panel \(\rightarrow\) Aggregated Diagnostic Report \(\rightarrow\) Coach Decision \(\rightarrow\) Next Observation
\end{quote}

This design corresponds to Equations~\ref{eq:coach-agent}--\ref{eq:coach-decision}: the judge panel observes the trajectory up to the current turn, the coach conditions on the panel's diagnostic report and Ada's current action, and the resulting decision changes the next observation only when intervention occurs.

The runtime coach and judges use DeepSeek-V3 in this condition. This is separate from the post-hoc automatic adjudication lens discussed in Appendix~\ref{app:llm-eval-align}; that evaluation layer uses a separate GLM, MiniMax, and Kimi combination and is not the mechanism that intervenes during Ada's trajectory. Across both the runtime Coach-Judge layer and the post-hoc adjudication layer, the judge models ran with model reasoning mode turned off, a temperature of 0.2, and an output limit of 2048 tokens. The reasoning mode was disabled because the judges were used as bounded assessment components under time and budget constraints rather than as open-ended deliberators.

\subsubsection{Judge Panel Composition}

The judge panel has six possible roles. Two roles are always active because they monitor the overall trajectory and the agent's reasoning stance. The other roles are activated only when the current turn falls within their domain. This keeps each turn evaluated by broad trajectory monitors plus the relevant action-specific monitor.

The upstream reason for this composition is the separation of concerns. \emph{Overall and bias} judges watch the trajectory of health and reasoning stance across all turns. \emph{Navigation}, \emph{cognitive}, \emph{conclusion}, and \emph{error} judges each inspect the local type of work being attempted. The coach, therefore, receives a report that combines persistent process-level monitoring with action-specific diagnosis.

\subsubsection{Judge Prompt Structure}

\begin{figure}[t]
\centering
\includegraphics[width=\textwidth]{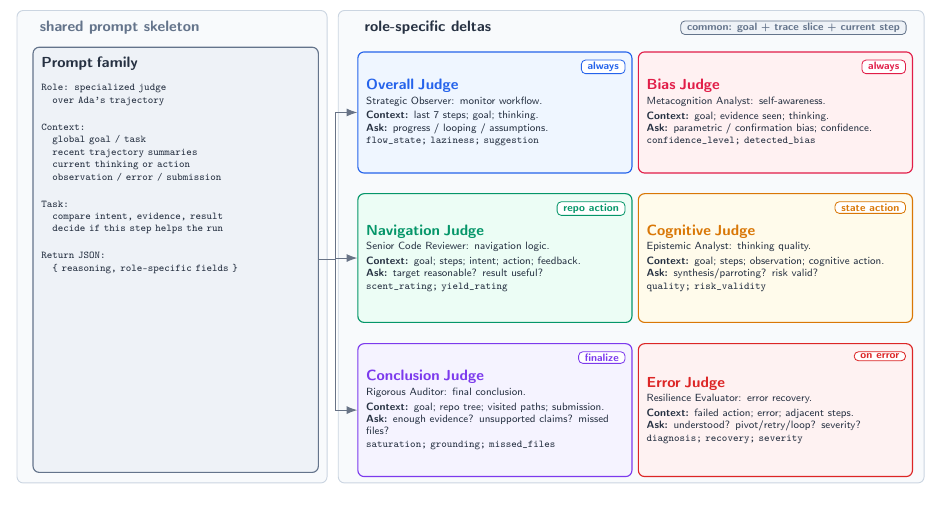}
\caption{Demonstration of the runtime judge prompts under the Coach-Judge condition. The prompt family shares a common skeleton over goal, trajectory slice, current step, and JSON reasoning output, while each judge specializes in the context slice and requested judgment for workflow progress, bias, navigation, cognitive state, conclusion, or error recovery.}
\label{fig:appendix-judge-prompts}
\end{figure}

The judge prompts share a common structure. A system-facing portion defines the judge's role and rating dimensions. A task-facing portion injects the current goal, a compact trajectory history, the current thinking block, the current action, and any action-specific context. The judge returns a structured assessment with a short rationale and categorical ratings. The categories differ by judge's role, as summarized in Figure~\ref{fig:appendix-judge-prompts}.

Before these judges run, each completed turn is summarized into a compact record of what Ada was trying to do, what action it took, and what information came back. The summaries form a rolling trajectory memory for the judge panel. This design avoids requiring every judge to reread the entire raw trajectory at every turn while still preserving recent context.

\subsubsection{Aggregation Logic}

The judge assessments are not averaged into a numerical score. Each judge contributes a short rationale and categorical ratings. The aggregation step converts these role-specific assessments into a diagnostic report for the coach. The report preserves the identity of each judge's role, its evaluation dimensions, and its reasoning. In effect, aggregation is a structured handoff from specialized observers to the coach, rather than a vote or scalar threshold.

This matters experimentally because intervention is not triggered by a single universal metric. The coach receives a distribution of diagnostic signals. For example, a weak navigation signal can be read differently if the overall judge still sees progress, while a premature conclusion signal has a stronger implication for rejection. The aggregation format, therefore, supports the coach's qualitative decision among \texttt{PASS}, \texttt{HINT}, and \texttt{REJECT}.

\subsubsection{Coach Decision Policy}

The coach is instructed to balance autonomy and rigor. Its default stance is non-intervention: if the judges report no serious problem, the trajectory should continue without alteration. The coach intervenes when multiple diagnostic signals indicate a harmful pattern, and it rejects conclusion attempts when the conclusion appears premature or unsupported.

{\small
\begin{longtable}{@{}>{\raggedright\arraybackslash}p{0.14\textwidth}>{\raggedright\arraybackslash}p{0.40\textwidth}>{\raggedright\arraybackslash}p{0.36\textwidth}@{}}
\caption{Coach decision policy in the intervention condition.}\label{tab:appendix-coach-decisions}\\
\toprule
Decision & Experimental meaning & Effect on Ada's next observation \\
\midrule
\endfirsthead
\toprule
Decision & Experimental meaning & Effect on Ada's next observation \\
\midrule
\endhead
\bottomrule
\endfoot
\texttt{PASS} & The current turn does not warrant intervention. Ada's autonomy is preserved. & The next observation contains only the normal environment response. \\
\texttt{HINT} & The trajectory shows a concerning but recoverable pattern, such as weak navigation plus stalled progress. & A concise coach message is appended as guidance. Ada may use or ignore it in the next turn. \\
\texttt{REJECT} & Ada attempts to close with a conclusion judged premature or insufficiently grounded. & The coach message is returned as a rejection of the attempted closure, and the trajectory continues. \\
\end{longtable}
}

The coach message is visible to Ada only under \texttt{HINT} or \texttt{REJECT}. Under \texttt{PASS}, the Coach-Judge process observes the trajectory but does not alter it. The condition therefore changes the content of some observations, not the formal turn structure of the trajectory.

\section{Appendix: Pilot Launches}\label{app:pilot-materials}

Before scaling to the full experimental matrix, three pilot launches were conducted to choose the prompt framework and to check that Ada could produce finite, recordable code-understanding trajectories under the seven-tool interface described in Appendix~\ref{app:experiment-pipeline}. The pilot launches used the same repository-level task and the same foundation model, GPT-4o, while varying only the surrounding prompt framework.

\subsection{Prompt Framework Selection}

The pilot compared three prompt-framework sources: a Copilot-style prompt from a public leaked-prompt collection, a Trae-style prompt from a public leaked-prompt collection, and a Cline-style prompt derived from an officially released open-source project. The pilot was not used to compare final answer quality. Its purpose was narrower: to determine whether each framework could sustain Ada's tool-mediated trajectory, reach explicit termination, and produce a usable code-understanding account.

\begin{figure}[t]
    \centering
    \includegraphics[width=0.65\linewidth]{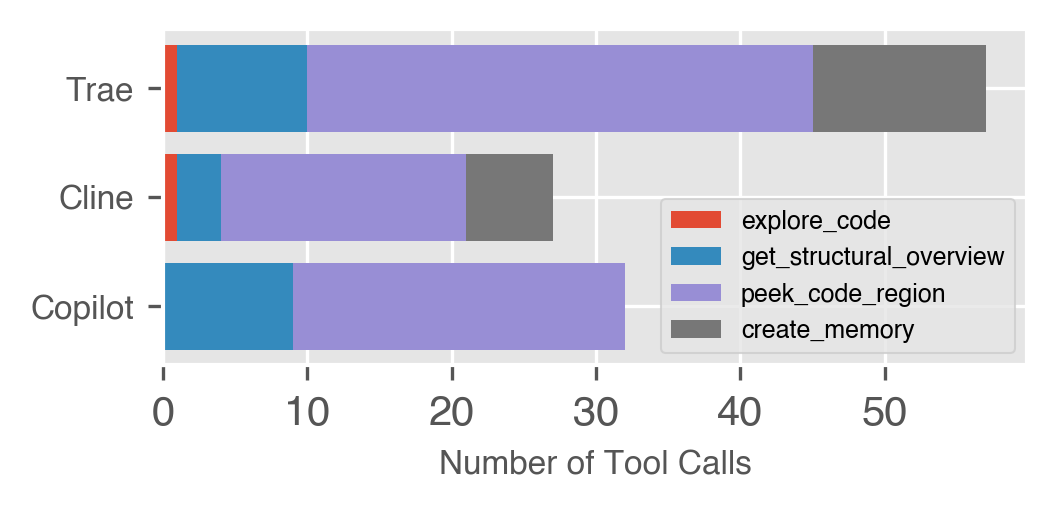}
    \caption{Pilot tool-call distributions across the three prompt frameworks. All three launches used the same task, the same foundation model, and the same seven-tool Ada interface. The figure reports the number of tool calls by action type, showing that all frameworks produced finite trajectories while differing in exploration volume and cognitive-tool use.}
    \label{fig:appendix-pilot-frameworks}
\end{figure}

All three preserved pilot trajectories reached an explicit conclusion action. The preserved records show different exploration volumes: Cline terminated after 28 recorded tool turns, Copilot after 33, and Trae after 58. The corresponding tool-call distributions are shown in Figure~\ref{fig:appendix-pilot-frameworks}. The pilot therefore established finite, recordable trajectories, while the stricter 50-turn bound was adopted later for the main matrix. Cline was selected for the main experiment because it avoided reliance on leaked prompt material, produced the shortest preserved pilot while still retaining enough exploration detail, and used a broader action mix than Copilot: four plotted tool types plus the terminal conclusion action, compared with Copilot's two plotted tool types plus the terminal conclusion action.

{\small
\begin{longtable}{@{}>{\raggedright\arraybackslash}p{0.12\textwidth}>{\raggedright\arraybackslash}p{0.18\textwidth}>{\raggedright\arraybackslash}p{0.60\textwidth}@{}}
\caption{Pilot framework comparison.}\label{tab:appendix-pilot-frameworks}\\
\toprule
Framework & Source status  & Role in design decision \\
\midrule
\endfirsthead
\toprule
Framework & Source status & Role in design decision \\
\midrule
\endhead
\bottomrule
\endfoot
Copilot & Public leaked-prompt collection from late 2024. &  Demonstrated that a short framework could sustain the tool protocol, but the preserved pilot used a narrower action mix, and the source was leaked material. \\
Trae & Public leaked-prompt collection from early 2025. &  Demonstrated that the tool protocol could remain stable in a longer pilot, but was not selected because the source was leaked material and the trajectory was the longest of the three. \\
Cline & Officially released open-source project. &  Selected for the main experiment because it used an official source, had a detailed prompt structure, and produced the shortest preserved pilot while still using navigation, memory, and traversal actions before termination. \\
\end{longtable}
}

\subsection{GLM Exclusion}

The pilot phase also tested the GLM family as a possible trajectory-generation model. GLM was excluded from the main trajectory-generation matrix because its tool-call compliance was not reliable enough for recoverable trajectory recording. Observed failure modes included malformed tool calls, calls to tools outside Ada's seven-tool interface, and loss of the required tool-call format within a trajectory. These failures made the affected pilot trajectories unsuitable as empirical objects for the observation lenses.

The exclusion applies only to trajectory generation. GLM was retained in the post-hoc automatic adjudication layer, where it contributes as one of the judge models described in Appendix~\ref{app:llm-eval-align}. In that role, GLM does not need to sustain Ada's multi-turn tool-use protocol; it evaluates completed trajectory materials under a separate adjudication design.

\section{LLM--Human Alignment and Adjudication Calibration}\label{app:llm-eval-align}

This appendix documents the post-hoc adjudication lens used in Section~\ref{sec:lenses}. The purpose is not to treat the automatic judges as a human substitute. The purpose is to make their protocol, internal agreement, human alignment, and length robustness visible enough that the Lens~3 results can be read with the right calibration.

\subsection{Evaluation Protocol and Judge Prompts}

The adjudication layer evaluates completed Ada trajectories through five evaluation families. Task Relevance, Trace Grounding, and Termination Calibration operate at the conclusion level: they ask whether the final conclusion addresses the task, whether it is grounded in the observed trajectory, and whether the stopping point was well calibrated. Navigation Prior Usage and Reasoning Prior Usage operate at the Turn or artifact level: Navigation Prior Usage evaluates whether navigation moves are led by evidence already visible in the trajectory, while Reasoning Prior Usage evaluates whether cognitive artifacts are organized by trajectory evidence or by prior-shaped framing.

The three conclusion-level evaluation families use a three-model vote over GLM-5, Kimi-K2.5, and MiniMax-M2.7. The two Prior Usage families use a single judge model. Evaluation items are anonymized with respect to the trajectory-generating model and experimental condition, and judge models are selected to avoid direct overlap with trajectory-generation models where possible. Each conclusion-level family contains 407 usable judged items, composed of 383 main-grid items and 24 Coach-Judge items; one additional candidate item was excluded after repeated server-side rejection. The Prior Usage evaluations contain 7,193 Navigation Prior Usage judgments and 929 Reasoning Prior Usage judgments. These judge runs used the same runtime settings as the Coach-Judge layer: model reasoning mode was turned off, temperature was fixed at 0.2, and outputs were capped at 2048 tokens.

\begin{figure}[t]
  \centering
  \includegraphics[width=\linewidth]{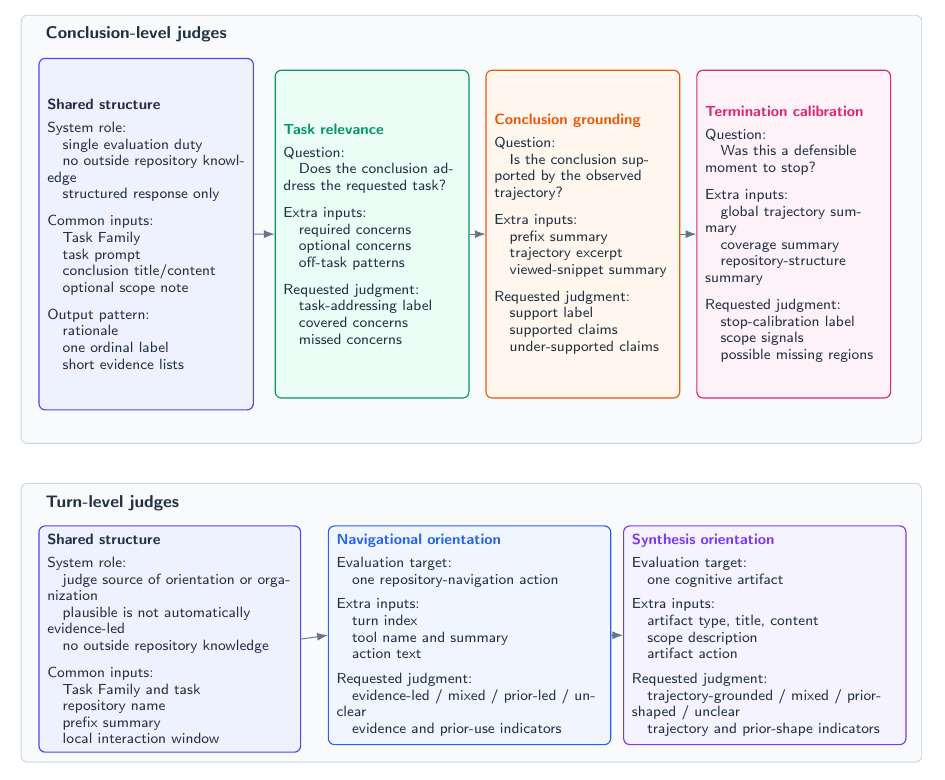}
  \caption{Structural templates for the post-hoc judge prompts. The diagrams show the shared input scaffold, the per-family evaluation target, and the requested output shape. They omit full prompt wording; the released prompt templates provide the exact wording for replication.}
  \label{fig:appendix-judge-prompt-structure}
\end{figure}

The three conclusion-level evaluations use four ordered labels, with occasional unclear or review flags handled outside the ordinal scale. The Prior Usage evaluations use three ordered evidence-orientation labels plus unclear: evidence-led or grounded maps to 0.0, mixed maps to 0.5, and prior-led or prior-shaped maps to 1.0 in the prior-leaning index. Figure~\ref{fig:appendix-judge-prompt-structure} summarizes the prompt structure without reproducing full prompt text.

\subsection{Inter-Judge Agreement}

Table~\ref{tab:appendix-judge-agreement} reports agreement among the current three conclusion-level judges. Task Relevance and Trace Grounding have moderate internal agreement. Termination Calibration is the reliability bottleneck: pairwise exact agreement is lower, ordinal disagreement is larger, and all three judges select the same label on only about one fifth of items.

\begin{table}[t]
\centering
\caption{Three-judge agreement for conclusion-level adjudication. The highlighted row marks the weakest agreement surface.}
\label{tab:appendix-judge-agreement}
\begin{tabular}{@{}lrrrr@{}}
\toprule
Evaluation family & \(n\) & Pair exact & Mean step diff. & All-same rate \\
\midrule
Task Relevance & 407 & 0.607 & 0.483 & 0.452 \\
Trace Grounding & 407 & 0.622 & 0.387 & 0.435 \\
\textbf{Termination Calibration} & \textbf{407} & \textbf{0.440} & \textbf{0.645} & \textbf{0.204} \\
\bottomrule
\end{tabular}
\end{table}

The lower reliability of Termination Calibration is partly explained by severity differences. Table~\ref{tab:appendix-t4-severity} reports a broader judge-severity audit for Termination Calibration. MiniMax-M2.7 is much more conservative than the other judges, with a mean Termination Calibration score of 2.631 on the 1--4 scale. Termination Calibration should therefore be read as a projected stopping-maturity signal, not as a mechanically obvious endpoint metric.

\begin{table}[t]
\centering
\caption{Judge severity audit for Termination Calibration. The highlighted row marks the severity outlier.}
\label{tab:appendix-t4-severity}
\begin{tabular}{@{}lr@{}}
\toprule
Judge & Mean Termination Calibration score \\
\midrule
DeepSeek-V3.2 & 3.646 \\
Kimi-K2.5 & 3.292 \\
Qwen3.5-397B-A17B & 3.209 \\
GLM-5 & 3.172 \\
\textbf{MiniMax-M2.7} & \textbf{2.631} \\
\bottomrule
\end{tabular}
\end{table}

\subsection{Human Alignment Check}

A separate human-alignment check sampled 250 items, with 50 items per evaluation family. One annotator supplied labels under the same information regime used for the automatic judges. Annotation notes report approximately 67 minutes of labeling decisions for about 150 items, which is itself informative: some dimensions, especially Trace Grounding, require cross-checking the conclusion against a full trajectory and are difficult under bounded human annotation time.

Table~\ref{tab:appendix-human-alignment} compares the sampled human labels against the LLM aggregate labels or scores. Task Relevance is the strongest human-aligned surface. Termination Calibration and Navigation Prior Usage are directionally useful but noisy. Trace Grounding and Reasoning Prior Usage are the weakest human-mirroring surfaces in this sample.

\begin{table}[t]
\centering
\caption{Human--LLM alignment by evaluation family. Task Relevance is highlighted as the strongest alignment; Trace Grounding and Reasoning Prior Usage are italicized as the weakest human-mirroring surfaces.}
\label{tab:appendix-human-alignment}
{\small
\setlength{\tabcolsep}{3.5pt}
\renewcommand{\arraystretch}{0.95}
\resizebox{\linewidth}{!}{%
\begin{tabular}{@{}lrrrrrrr@{}}
\toprule
Eval & \(n\) & Pearson \(r\) & Spearman \(\rho\) & Weighted \(\kappa\) & Exact & Within~1 & Mean diff. \\
\midrule
\textbf{Task Relevance} & \textbf{50} & \textbf{0.608} & \textbf{0.616} & \textbf{0.547} & \textbf{44\%} & \textbf{84\%} & \textbf{+0.252} \\
\emph{Trace Grounding} & \emph{50} & \emph{-0.135} & \emph{-0.089} & \emph{0.116} & \emph{54\%} & \emph{98\%} & \emph{+0.264} \\
Termination Calibration & 50 & 0.318 & 0.297 & 0.237 & 40\% & 88\% & -0.068 \\
Navigation Prior Usage & 50 & 0.282 & 0.312 & 0.305 & 48\% & 80\% & -0.240 \\
\emph{Reasoning Prior Usage} & \emph{50} & \emph{0.077} & \emph{0.150} & \emph{0.106} & \emph{42\%} & \emph{80\%} & \emph{+0.160} \\
\bottomrule
\end{tabular}%
}
}
\end{table}

Table~\ref{tab:appendix-model-human-gap} compares model--model alignment with human--model alignment on the three conclusion-level families. The Trace Grounding gap is the clearest warning sign: judge models agree with one another more than they agree with the human annotator.

\begin{table}[t]
\centering
\caption{Model--model alignment versus human--model alignment for the three conclusion-level families. The bold Trace Grounding human-model values mark the critical alignment gap.}
\label{tab:appendix-model-human-gap}
{\small
\setlength{\tabcolsep}{3.5pt}
\renewcommand{\arraystretch}{0.95}
\resizebox{\linewidth}{!}{%
\begin{tabular}{@{}lrrrr@{}}
\toprule
Eval & Model--model \(r\) & Human--model \(r\) & Model--model \(\kappa\) & Human--model \(\kappa\) \\
\midrule
Task Relevance & 0.854 & 0.571 & 0.839 & 0.520 \\
Trace Grounding & 0.318 & \textbf{-0.089} & 0.269 & \textbf{-0.077} \\
Termination Calibration & 0.458 & 0.243 & 0.396 & 0.224 \\
\bottomrule
\end{tabular}%
}
}
\end{table}

\subsection{Interpretation}

Task Relevance is the most trustworthy automatic adjudication surface in the current validation. Its Pearson correlation with the human labels is 0.608, and its weighted \(\kappa\) is 0.547. Main-text claims that rely on whether a conclusion addressed the task therefore rest on the firmest Lens~3 ground.

Trace Grounding and Reasoning Prior Usage should be read differently. They are machine-consensus projections, not human proxies. The judges can process the whole trajectory without fatigue, while a bounded human annotation session must cross-reference conclusions, intermediate observations, and prior-shaped reasoning under time pressure. The present data cannot distinguish whether model agreement reflects shared model priors or better large-context cross-referencing. What it establishes is more limited and more useful: Trace Grounding and Reasoning Prior Usage gain credibility when they converge with other observation lenses, not when treated as standalone human-aligned measures.

Termination Calibration and Navigation Prior Usage occupy a middle ground. Their correlations with human labels are positive, but weak enough that the paper should lean on label-composition patterns rather than small mean-score differences. For Termination Calibration, this means reading shifts such as ready-to-finalize versus borderline as qualitative stopping profiles. For Navigation Prior Usage, it means using the prior-leaning index as a directional signal about navigation orientation, not as a precise measure of hidden intent.

The alignment check also reveals an asymmetry in evaluation difficulty. Dimensions that compare a conclusion against the task are more human-accessible than dimensions that require checking a conclusion against the full trajectory or inferring prior familiarity from behavior. This is not only a limitation of the validation exercise; it is also a result about which evaluation questions are tractable for human annotation at scale.

\subsection{Length Robustness Checks}

Table~\ref{tab:appendix-length-robustness} reports the conclusion-length robustness check for the main non-coach grid. Raw correlations are non-trivial, especially for Task Relevance and Trace Grounding, but the added explanatory power after controlling for model, condition, Task Family, and Repository is small. The corresponding visual diagnostic for conclusion length and Task Relevance appears in Figure~\ref{fig:appendix-conclusion-length}. Table~\ref{tab:appendix-t4-length} reports the analogous trajectory-length check for Termination Calibration.

\begin{table}[t]
\centering
\caption{Conclusion length versus conclusion-level adjudication scores, main non-coach grid.}
\label{tab:appendix-length-robustness}
\begin{tabular}{@{}lrrr@{}}
\toprule
Evaluation family & \(n\) & Raw Pearson \(r\) & Added \(\Delta R^2\) \\
\midrule
Task Relevance & 383 & 0.533 & 0.016 \\
Trace Grounding & 383 & -0.316 & 0.004 \\
Termination Calibration & 383 & 0.300 & 0.001 \\
\bottomrule
\end{tabular}
\end{table}

\begin{table}[t]
\centering
\caption{Trajectory length versus Termination Calibration, main non-coach grid.}
\label{tab:appendix-t4-length}
\begin{tabular}{@{}lrrrr@{}}
\toprule
Metric & \(n\) & Raw Pearson \(r\) & Raw \(R^2\) & Added \(\Delta R^2\) \\
\midrule
Valid tool turns & 383 & 0.258 & 0.067 & 0.002 \\
\bottomrule
\end{tabular}
\end{table}

The model-level length table explains why raw correlations appear in the first place. Conclusion length is strongly entangled with the trajectory-generating model. Qwen3.5-397B and DeepSeek-V3.2 produce much longer conclusions than GPT-4.1 and GPT-4o, so a raw length association can partly reflect model identity rather than an independent length effect.

\begin{table}[t]
\centering
\caption{Conclusion length by trajectory-generating model, main non-coach grid.}
\label{tab:appendix-length-by-model}
\begin{tabular}{@{}lrrr@{}}
\toprule
Model & \(n\) & Mean words & Median words \\
\midrule
Qwen3.5-397B & 95 & 323.1 & 311.0 \\
DeepSeek-V3.2 & 96 & 254.7 & 246.5 \\
GPT-4o & 96 & 125.0 & 124.0 \\
GPT-4.1 & 96 & 97.0 & 89.5 \\
\bottomrule
\end{tabular}
\end{table}

Length is therefore a visible proxy for answer development, but not an independent driver of judge scores in the controlled checks. The automatic adjudication lens should be read as a set of semantic projections conditioned by model, task, condition, and repository, rather than as a disguised length measure.

\section{Supplementary Results and Unreported Details}\label{app:unreported-details}

This appendix reports selected supplementary materials that support claims compressed in Section~\ref{sec:results}. The materials are organized by observation lens and are included only where they clarify a specific claim, caveat, or extension.

\subsection{Tool-Use Concentration and Task Variation}

\begin{figure}[!h]
  \centering
  \begin{subfigure}[b]{0.48\linewidth}
    \centering
    \includegraphics[width=\linewidth]{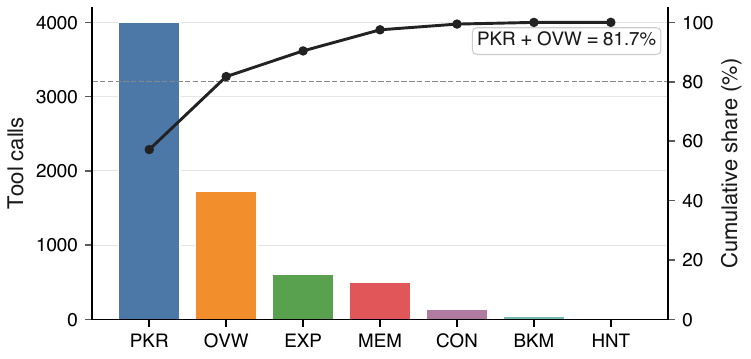}
    \caption{Overall concentration.}
    \label{fig:appendix-tool-distribution-a}
  \end{subfigure}\hfill
  \begin{subfigure}[b]{0.48\linewidth}
    \centering
    \includegraphics[width=\linewidth]{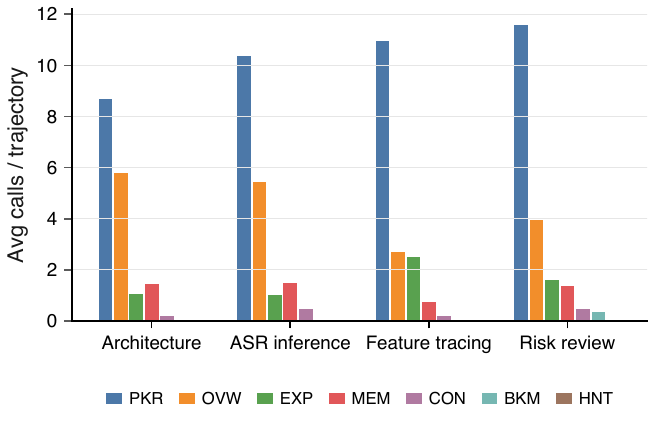}
    \caption{Variation by Task Family.}
    \label{fig:appendix-tool-distribution-b}
  \end{subfigure}
  \caption{Supplementary surface view of Ada's tool use. Panel~(\subref{fig:appendix-tool-distribution-a}) shows tool-call concentration across the full main matrix. Panel~(\subref{fig:appendix-tool-distribution-b}) shows average tool-call composition by Task Family, indicating that task openings modulate the balance of structural orientation, close reading, traversal, and cognitive actions.}
  \label{fig:appendix-tool-distribution}
\end{figure}

The surface activity view shows that Ada's visible tool use is highly concentrated. Figure~\ref{fig:appendix-tool-distribution} reports the overall tool-call distribution and its variation by Task Family. Across the main matrix, close code reading and structural overview account for most actions. Task Family changes the balance of those actions: feature-implementation tasks increase close reading, while broader repository-understanding openings retain more structural orientation.


\subsection{Extended Action-Policy Projections}\label{appsub:extended-action-policy}

\begin{figure}[t]
  \centering
  \includegraphics[width=0.8\linewidth]{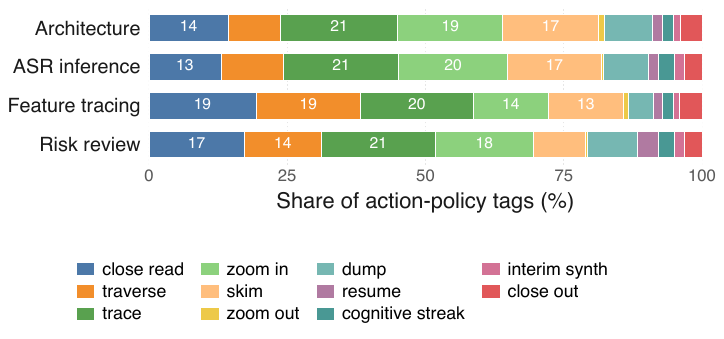}
  \caption{Task Family changes the transition mix even under the same apparatus. Stacked bars report the share of adjacent-action transition tags, including close read, traverse, trace, zoom in, skim, dump, cognitive streak, interim synthesis, and close out.}
  \label{fig:appendix-action-policy}
\end{figure}

Figure~\ref{fig:appendix-action-policy} extends the action-policy lens used in the main text. The Combined Condition is included here because the main Results figure omits it for compactness. Across models, Combined Condition profiles closely follow the pressure-family pattern rather than the Natural or Prior Suppression pattern. Task Family also shifts action policy: Feature Implementation Tracing increases close read and traverse shares, while Architecture Analysis and ASR Inference retain more trace and skim behavior.

\begin{figure}[t]
  \centering
  \includegraphics[width=0.88\linewidth]{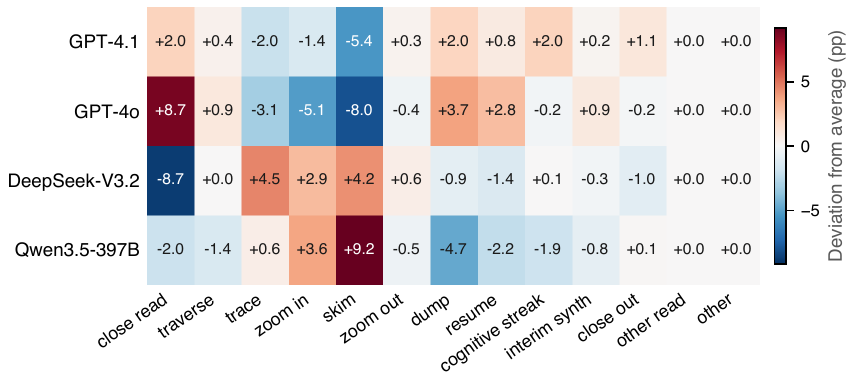}
  \caption{Model-level deviations from the average action-policy profile. Positive values indicate that a model uses a transition tag more than the cross-model average, and negative values indicate lower-than-average use.}
  \label{fig:appendix-model-deviation}
\end{figure}

The model-deviation heatmap in Figure~\ref{fig:appendix-model-deviation} gives a relative view of the same action-policy lens. It centers each transition tag against the cross-model average, making model-characteristic signatures visible: GPT-4o is close-read heavy, DeepSeek-V3.2 is trace-heavy, and Qwen3.5-397B is skim-heavy. This view supplements the stacked bars by showing deviation from the common baseline rather than absolute composition.

\begin{figure}[t]
  \centering
  \includegraphics[width=0.88\linewidth]{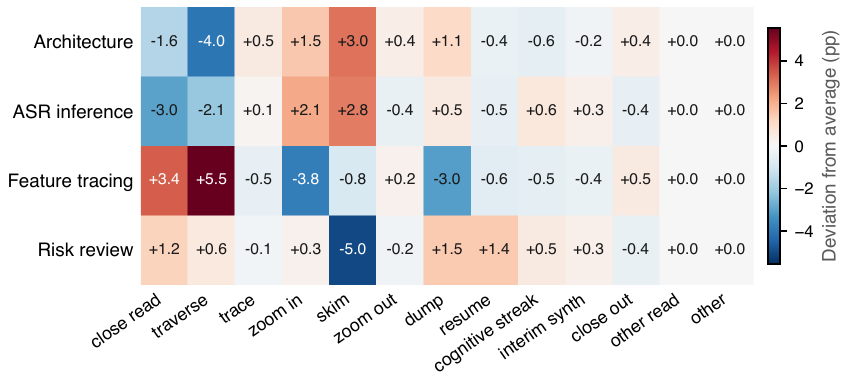}
  \caption{Task-Family deviations from the average action-policy profile. Positive values indicate that a Task Family uses a transition tag more than the cross-task average, and negative values indicate lower-than-average use. This figure supplies the relative-to-average task view omitted from the main text.}
  \label{fig:appendix-task-deviation}
\end{figure}

Figure~\ref{fig:appendix-task-deviation} adds the matching relative-to-average view for Task Family. Feature Implementation Tracing over-indexes on close read and traverse while under-indexing on skim, making its implementation-following stance visible in a single panel. Architecture Analysis and ASR Inference tilt more toward skim and zoom-in behavior, while Risk and Code-Smell Review carries the clearest interim-synthesis and dump deviations. This heatmap supports the same claim as Figure~\ref{fig:appendix-action-policy}, but in deviation form rather than composition form.

\subsection{Supplementary Adjudication Diagnostic}\label{appsub:supplementary-adjudication-diagnostic}

\begin{figure}[t]
  \centering
  \includegraphics[width=0.70\linewidth]{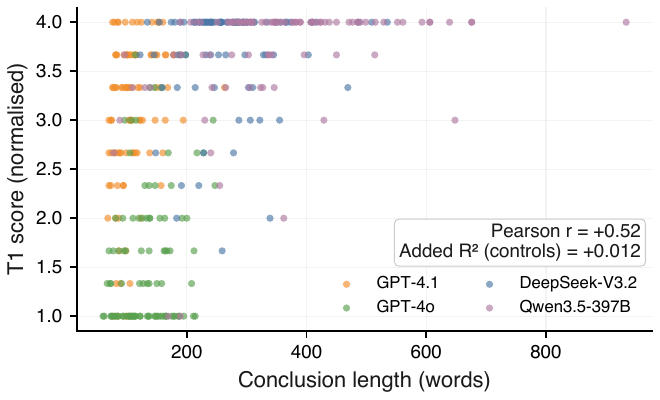}
  \caption{Conclusion length versus Task Relevance. The pooled association is positive, but the diagnostic should be read cautiously: length may correlate with fuller task coverage without guaranteeing stronger grounding.}
  \label{fig:appendix-conclusion-length}
\end{figure}

{\small
\begin{longtable}{@{}>{\raggedright\arraybackslash}p{0.20\linewidth}rrr>{\raggedleft\arraybackslash}p{0.18\linewidth}@{}}
\caption{Conclusion-length diagnostic by model. Task Relevance is reported on the original 1--4 adjudication scale.}\label{tab:appendix-conclusion-length}\\
\toprule
Model & Traj. & Median words & Mean Task Relevance & Length--Task Relevance $r$ \\
\midrule
\endfirsthead
\toprule
Model & Traj. & Median words & Mean Task Relevance & Length--Task Relevance $r$ \\
\midrule
\endhead
\bottomrule
\endfoot
DeepSeek-V3.2 & 96 & 262.0 & 3.719 & 0.115 \\
GPT-4.1 & 96 & 102.5 & 3.035 & 0.305 \\
GPT-4o & 96 & 134.0 & 1.597 & 0.230 \\
Qwen3.5-397B & 96 & 341.0 & 3.719 & 0.372 \\
\end{longtable}
}

Conclusion length is reported here as a diagnostic, not as a quality measure. Longer conclusions can appear more complete, but length does not by itself establish repository grounding. Figure~\ref{fig:appendix-conclusion-length} shows a moderate positive association between conclusion length and Task Relevance in the pooled data, while the added explanatory value after controls is small. Table~\ref{tab:appendix-conclusion-length} reports the compact model-level summary.

Overall, the pooled length--Task Relevance correlation is \(r=0.521\), while the controlled added \(R^2\) reported in the diagnostic figure is \(0.012\). The appendix therefore treats length as a cautionary surface diagnostic rather than as a substitute for the adjudication lens.

\subsection{Repository-Footprint Anchors and Transition Corridors}

{\scriptsize
\begin{longtable}{@{}>{\raggedright\arraybackslash}p{0.14\linewidth}>{\raggedright\arraybackslash}p{0.76\linewidth}@{}}
\caption{Recurring anchor files by Repository, ranked by average position across conditions.}\label{tab:appendix-anchor-files}\\
\toprule
Repository & Top recurring anchors \\
\midrule
\endfirsthead
\toprule
Repository & Top recurring anchors \\
\midrule
\endhead
\bottomrule
\endfoot
Flask & \texttt{src/flask/app.py}; \texttt{src/flask/sansio/app.py}; \texttt{src/flask/sansio/scaffold.py} \\
Graphiti & \texttt{graphiti\_core/graphiti.py}; \texttt{graphiti\_core/search/search.py}; \texttt{graphiti\_core/edges.py} \\
Prompt Optimizer & \texttt{middleware.js}; \texttt{packages/core/src/services/llm/service.ts}; \texttt{packages/core/src/services/prompt/service.ts} \\
Requests & \texttt{src/requests/sessions.py}; \texttt{src/requests/\_\_init\_\_.py}; \texttt{src/requests/adapters.py} \\
Scikit-learn & \texttt{sklearn/tree/\_classes.py}; \texttt{sklearn/base.py}; \texttt{sklearn/tree/\_tree.pyx} \\
TensorZero & \texttt{tensorzero-core/src/endpoints/inference.rs}; \texttt{gateway/src/main.rs}; \texttt{tensorzero-core/src/config\_parser.rs} \\
\end{longtable}
}

{\scriptsize
\begin{longtable}{@{}>{\raggedright\arraybackslash}p{0.13\linewidth}>{\raggedright\arraybackslash}p{0.56\linewidth}>{\raggedleft\arraybackslash}p{0.10\linewidth}>{\raggedleft\arraybackslash}p{0.11\linewidth}@{}}
\caption{Top file-to-file transition corridors by Repository.}\label{tab:appendix-transition-corridors}\\
\toprule
Repository & Corridor & Moves & Traj. \\
\midrule
\endfirsthead
\toprule
Repository & Corridor & Moves & Traj. \\
\midrule
\endhead
\bottomrule
\endfoot
Flask & \texttt{src/flask/app.py} \(\rightarrow\) \texttt{src/flask/sansio/app.py} & 20 & 17 \\
Graphiti & \texttt{graphiti\_core/graphiti.py} \(\rightarrow\) \texttt{graphiti\_core/search/search.py} & 12 & 11 \\
Prompt Optimizer & \texttt{packages/core/src/services/prompt/service.ts} \(\rightarrow\) \texttt{packages/core/src/services/template/processor.ts} & 9 & 9 \\
Requests & \texttt{src/requests/sessions.py} \(\rightarrow\) \texttt{src/requests/adapters.py} & 16 & 14 \\
Scikit-learn & \texttt{sklearn/\_\_init\_\_.py} \(\rightarrow\) \texttt{sklearn/base.py} & 8 & 8 \\
TensorZero & \texttt{tensorzero-core/src/lib.rs} \(\rightarrow\) \texttt{tensorzero-core/src/endpoints/mod.rs} & 10 & 10 \\
\end{longtable}
}

The footprint lens can be summarized without showing another full heatmap. Table~\ref{tab:appendix-anchor-files} reports recurring anchor files: files that repeatedly appear among the most visited files for a repository across conditions. Table~\ref{tab:appendix-transition-corridors} reports the most frequent file-to-file movement corridors. These tables expose recurring spatial attention in the observed trajectories; they do not claim that the files are objectively the most important files in the repository.

\FloatBarrier

\subsection{Coach-Judge Condition Details}\label{appsub:coach-judge-condition-details}

\FloatBarrier

{\small
\begin{longtable}{@{}
  >{\raggedright\arraybackslash}p{0.20\linewidth}
  >{\raggedleft\arraybackslash}p{0.17\linewidth}
  >{\raggedleft\arraybackslash}p{0.15\linewidth}
  >{\raggedleft\arraybackslash}p{0.19\linewidth}
  >{\raggedleft\arraybackslash}p{0.15\linewidth}
@{}}
\caption{Paired Natural versus Coach-Judge surface/action statistics for GPT-4.1.}\label{tab:appendix-coach-paired}\\
\toprule
Metric & Natural mean & Coach mean & Mean difference & Sign-test $p$ \\
\midrule
\endfirsthead
\toprule
Metric & Natural mean & Coach mean & Mean difference & Sign-test $p$ \\
\midrule
\endhead
\bottomrule
\endfoot
Turns & 18.8 & 48.1 & +29.3 & $<.001$ \\
Elapsed time (s) & 94.4 & 286.2 & +191.8 & .0015 \\
Conclusion attempts & 0.5 & 8.3 & +7.8 & $<.001$ \\
Error turns & 0.3 & 4.6 & +4.3 & $<.001$ \\
Coach-message turns & 0.0 & 11.2 & +11.2 & $<.001$ \\
\end{longtable}
}

Table~\ref{tab:appendix-coach-paired} reports paired supplementary numbers for GPT-4.1 under Natural and Coach-Judge conditions. The table documents the scale of added interaction without repeating the full RQ3 interpretation. Coach-Judge trajectories contain more turns, more conclusion attempts, more error turns, and direct coach messages in a substantial share of turns.

\begin{table}[t]
\centering
\caption{Per-tool Turn shifts under Natural and Coach-Judge conditions for GPT-4.1. PKR = \texttt{peek\_code\_region}; OVW = \texttt{get\_structural\_overview}; EXP = \texttt{explore\_code}.}
\label{tab:appendix-coach-tool-shift}
\begin{tabular}{@{}lrrrr@{}}
\toprule
Tool & Natural mean & Coach mean & Mean difference & Sign-test \(p\) \\
\midrule
PKR & 10.58 & 19.08 & +8.50 & 0.0227 \\
OVW & 3.00 & 13.71 & +10.71 & \(<.001\) \\
EXP & 2.79 & 3.29 & +0.50 & 1.0000 \\
\bottomrule
\end{tabular}
\end{table}

Table~\ref{tab:appendix-coach-tool-shift} adds the missing per-tool breakdown for the same paired comparison. The largest absolute shift is in structural overview activity, while explore-code usage is nearly flat. The added interaction therefore looks less like a new traversal strategy and more like repeated re-orientation plus additional close reading.

{\small
\begin{table}[t]
\centering
\caption{Trace Grounding label distribution under Natural and Coach-Judge conditions.}
\label{tab:appendix-t2-labels}
\begin{tabular}{@{}>{\raggedright\arraybackslash}p{0.40\linewidth}rr@{}}
\toprule
Trace Grounding label & Natural & Coach-Judge \\
\midrule
Well supported & 10 (41.7\%) & 3 (12.5\%) \\
Partially supported & 14 (58.3\%) & 19 (79.2\%) \\
Thinly supported & 0 (0.0\%) & 2 (8.3\%) \\
\bottomrule
\end{tabular}
\end{table}
}

\FloatBarrier

The direct coach messages are split into 179 hint turns and 39 rejection turns among 269 coach-message turns. This distribution supports the main-text reading that the intervention condition often extended trajectories through guidance and conclusion rejection rather than through a change to the formal turn structure. At the tool level, the same pattern appears numerically: PKR rises from 10.58 to 19.08 mean turns, OVW rises from 3.00 to 13.71, and EXP changes only from 2.79 to 3.29.

Table~\ref{tab:appendix-t2-labels} reports the Trace Grounding label distribution for the same paired comparison. The Coach-Judge condition reduces the share of well-supported conclusions and introduces thinly supported cases. This table gives the categorical detail behind the main RQ3 adjudication panel.

\end{appendices}

\bibliography{sn-bibliograph}

\end{document}